\documentclass[runningheads]{llncs}

\usepackage{eccv}

\usepackage{eccvabbrv}

\usepackage{graphicx}

\usepackage[accsupp]{axessibility}  
\usepackage{booktabs}
\usepackage[table]{xcolor}
\usepackage{colortbl}
\usepackage{graphicx}
\usepackage{array}

\newcommand{\best}[1]{\textbf{#1}}
\newcolumntype{C}[1]{>{\centering\arraybackslash}p{#1}}

\usepackage[table]{xcolor}
\usepackage{makecell}

\usepackage{hyperref}

\usepackage{orcidlink}
\usepackage{float}

\begin{document}

\title{Next-Acceleration-Scale Prediction for Autoregressive MRI Reconstruction} 


\author{Yilmaz Korkmaz\inst{1} \and
Vishal M. Patel\inst{1}}

\authorrunning{Y. Korkmaz and V.M. Patel}

\institute{Johns Hopkins University, Baltimore MD 21218, USA\\
\email{\{ykorkma1,vpatel36\}@jhu.edu}}

\maketitle

\begin{abstract}
MRI reconstruction is an inherently ill-posed inverse problem, since incomplete measurements admit many plausible solutions. This ambiguity becomes more severe under high acceleration, where pixel-domain continuous predictors tend to average over feasible reconstructions and suppress high-frequency anatomy. We address this limitation by moving reconstruction to \textbf{discrete multi-scale latent space} and posing it as autoregressive next-acceleration-scale prediction. Leveraging discrete priors proven effective in visual autoregressive modeling, our method restricts the solution to compact sequences of codebook tokens, enabling sharp reconstructions even from extremely sparse measurements. This discrete autoregressive formulation also aligns naturally with modern large language model post-training techniques. Building on this observation, we introduce \textbf{on-policy privileged information distillation} for visual autoregressive modeling, where a teacher is provided training only privileged context that is unavailable at inference, in our case fully sampled acquisitions, and supervises a student trained on its own rollouts, leading to consistent reconstruction gains. Through extensive experiments on the fastMRI benchmark, we show that our approach delivers improved reconstruction performance across diverse sampling patterns under extreme undersampling. Project website is \href{https://yilmazkorkmaz1.github.io/discrete-mri-reconstruction-opd/}{here}. 

\end{abstract}

\begin{figure}
\centering
    \includegraphics[width=1\linewidth]{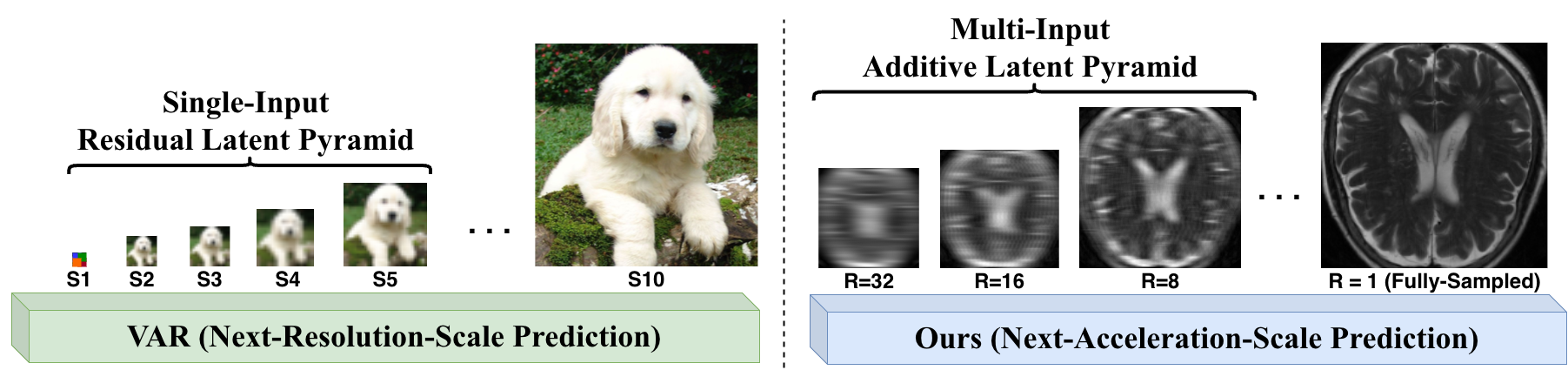}
    \caption{Left: VAR \cite{tian2024visual}  constructs a residual latent pyramid by progressively downsampling and quantizing the residual continuous latent of a single input image, generating content in a coarse-to-fine manner via next-resolution (next-scale) prediction. Right: In our method, the hierarchy is induced before encoding by applying MRI native Fourier undersampling at different acceleration factors (R), yielding a multi-input latent pyramid aligned with acceleration scales. Our scheme naturally enables next-acceleration-scale prediction to reconstruct the fully-sampled acquisition from undersampled inputs.}
    \label{fig:teaser}
\end{figure}

\section{Introduction}
Magnetic Resonance Imaging (MRI) provides excellent soft-tissue contrast, but long acquisition times remain a major practical limitation. MRI data are acquired in k-space, the frequency-domain representation of the image, and accelerated imaging reconstructs the final image from only a subset of these measurements ~\cite{lustig2007sparse}. Recent deep learning methods have significantly improved this reconstruction process, but under extreme acceleration the problem remains severely ill-posed, and current methods often recover global structure while failing to preserve diagnostically important high-frequency anatomy \cite{radmanesh2022exploring}. This limitation suggests that accurate reconstruction under severe undersampling may require a representation that constrains the solution space more strictly than direct pixel-level prediction.

Recent visual autoregressive modeling (VAR) offers such a perspective by showing that high-fidelity images can be generated from compact discrete latent sequences~\cite{tian2024visual}. By representing images as spatial grids of codebook indices, these models replace redundant pixel-level representations with discrete tokens that better capture structured visual content. We observe that this discrete latent formulation is also well suited to accelerated MRI, where faithful reconstruction depends not only on pixel fidelity but also on preserving anatomically coherent structure under severe ambiguity.

Building on this observation, we reformulate VAR for MRI reconstruction by replacing resolution-wise generation with prediction across acceleration levels within a discrete latent hierarchy. Rather than progressively refining spatial resolution, the model predicts finer reconstruction tokens conditioned on latent representations from higher acceleration factors, which provide anatomical context. To enable this formulation, we replace VAR’s residual latent hierarchy for a single input image with an additive multi-input hierarchy that jointly organizes latent representations from multiple acceleration levels (see \cref{fig:teaser}). The resulting discrete codebook serves as a learned vocabulary of plausible anatomical structures, imposing a strong inductive bias against over-smoothed or anatomically implausible reconstructions. We further adapt the autoregressive transformer to meet the stricter spatial fidelity demands of MRI reconstruction (\cref{sec:var_section}).

Our discrete formulation also aligns naturally with recent post-training advances in large language models built on next-token prediction. Leveraging this connection, we introduce an on-policy privileged information distillation strategy for visual autoregressive modeling. During training, a privileged teacher observes fully sampled acquisitions and supervises the student along the student’s own autoregressive rollouts, improving token prediction under imperfect contexts while leaving inference-time inputs unchanged. By guiding generation under ambiguous rollout states, this strategy reduces hallucinated anatomy and yields consistent reconstruction gains across diverse sampling patterns and MRI contrasts. Our contributions are summarized as follows:
\begin{itemize}
    \item We introduce a discrete autoregressive MRI reconstruction framework that casts accelerated MRI recovery as next-acceleration-scale prediction in a multi-scale latent token hierarchy, improving preservation of anatomically meaningful high-frequency detail under extreme undersampling.
    
    \item We design a tailored architecture for this framework, consisting of an additive multi-input vector-quantized variational autoencoder (AQ-VAE) with a shared discrete codebook across acceleration levels, together with a cross-attentive autoregressive transformer for high-fidelity next-scale token prediction.
    
    \item We propose, to the best of our knowledge, the first on-policy privileged information distillation method for VAR models, using training-only privileged context to supervise a student on its own rollouts without changing inference-time inputs.
\end{itemize}

\section{Related Work}

\subsection{Deep Learning Methods for MRI Reconstruction}
Deep learning has substantially advanced accelerated MRI reconstruction. Early approaches mainly relied on CNN-based architectures that learned image-domain priors and reconstruction mappings directly from data~\cite{wang2020neural,ChulYe2018,rgan,Hyun2018}. More recent work explored transformer-based models for improved long-range dependency modeling~\cite{guo2023reconformer,huang2022swin} and Mamba-based architectures as efficient alternatives for sequence modeling~\cite{zou2024mmr,kabas2024physics,korkmaz2025mambarecon}. In parallel, physics-guided methods explicitly incorporated the MRI forward model into trainable reconstruction pipelines, improving data consistency and robustness~\cite{MoDl,schlemper2017deep,yaman2020,yiasemis2022recurrent,Variatonal_end2end}. Diffusion-based approaches further expanded the space of reconstruction priors and demonstrated strong performance in challenging undersampling regimes~\cite{dar2022adaptive,peng2022towards,korkmaz2023self,zhang2025mdpg}. In contrast to these continuous reconstruction approaches, our method adopts a discrete autoregressive formulation that models reconstruction as structured token prediction across acceleration levels.

\subsection{From Discrete Tokenization to Visual Autoregressive Modeling}

Discrete latent modeling shifted autoregressive image generation from pixel sequences to compact visual token sequences. VQ-VAE introduced a tokenizer that maps an image to a lower-dimensional latent grid and quantizes each latent vector using a learned codebook~\cite{van2017neural}. VQ-GAN improved perceptual reconstruction quality through adversarial and perceptual objectives~\cite{esser2021taming}, while hierarchical and residual quantization schemes increased representational capacity without excessively large codebooks~\cite{razavi2019generating,lee2022autoregressive}. These developments enabled scalable autoregressive modeling over discrete visual tokens~\cite{van2017neural,razavi2019generating,esser2021taming,ramesh2021zero,lee2022autoregressive}.

VAR further reformulated autoregressive image generation as scale-wise prediction rather than raster-scan token prediction~\cite{tian2024visual}. By generating all tokens of a given scale in parallel and proceeding from coarse to fine resolutions, VAR reduces sequential decoding cost while maintaining strong image generation quality. Following VAR, several extensions have adapted visual autoregressive modeling to tasks such as segmentation~\cite{zheng2025seg}, image restoration~\cite{wang2025navigating,rajagopalan2025restorevar}, and conditional generation~\cite{li2024controlvar}. A smaller body of work has also explored related ideas in medical imaging, including medical image generation~\cite{he2026medvar}, synthetic data generation for federated MRI reconstruction~\cite{nezhad2025generative}, pathological image restoration~\cite{liu2025conditional}, and medical video segmentation~\cite{yao2025hrvvs}.

\subsection{On-Policy Information Distillation}
Most recent work has revisited on-policy distillation as a post-training strategy for large language models, leveraging self-generated rollouts to better align training with inference-time behavior and improve reasoning and agentic capabilities. In \cite{agarwal2024policy}, on-policy distillation trains student on its own rollouts and uses teacher feedback on those same rollouts to reduce distribution mismatch. In \cite{shenfeld2026self}, self-distillation is studied in continual learning using an EMA teacher to stabilize updates and mitigate forgetting. In \cite{zhao2026self}, on-policy self-distillation is applied to reasoning by training on self-generated solutions paired with improved targets. In \cite{hubotter2026reinforcement}, self-distillation is integrated into reinforcement learning-style optimization to improve policy learning stability. In \cite{penaloza2026privileged}, training-time privileged information is distilled through a joint teacher-student objective. 

Our method adapts this emerging post-training paradigm from large language models to visual autoregressive modeling. The student is trained on its own rollouts, while the teacher is provided with additional training-time privileged context that is unavailable at inference. Prior work with LLMs has used successful agentic trajectories~\cite{penaloza2026privileged}, ground-truth answers~\cite{zhao2026self}, or self-reflective feedback~\cite{hubotter2026reinforcement} as privileged information. In our case, the privileged context is the fully sampled MRI acquisition.

\section{Preliminaries}

\subsection{Accelerated MRI Reconstruction}
Accelerated MRI recovers the target image $x$ from undersampled k-space measurements $y_{\Omega}$ by inverting the encoding operator $E_{\Omega}$ (which combines coil sensitivities and the partial Fourier transform on the sampling set $\Omega$). This is typically achieved by minimizing a data-consistency term $\| E_{\Omega} x - y_{\Omega} \|_{2}^{2}$ regularized by a prior $\mathcal{R}(x)$. While conventional methods learn $\mathcal{R}(x)$ in the continuous pixel domain, we propose learning this prior in a discrete latent space.

\subsection{Next-Resolution-Scale Prediction (VAR)}

VAR models image generation as a hierarchical autoregressive process over discrete latents, where each finer-resolution latent is predicted from previously generated coarser ones~\cite{tian2024visual}. These latents are obtained from a single image by progressively quantizing residual latent components across resolutions. Let $\{Q_1,\dots,Q_S\}$ denote the multi-scale discrete latents, ordered from coarse to fine, with $Q_s^{\text{next}} = Q_{s+1}$ for $s=1,\dots,S-1$. The joint prior is factorized as
\begin{equation}
p_{\theta}(Q_1, \dots, Q_S)
=
\prod_{s=1}^{S-1}
p_{\theta}\!\left(
Q_{s+1}\mid Q_1, \dots, Q_s
\right),
\label{eq:var_scalewise_likelihood}
\end{equation}
with each conditional modeled by an autoregressive transformer.

\section{Next-Acceleration-Scale Prediction}
\label{sec:var_section}

We formulate accelerated MRI reconstruction as next-acceleration-scale prediction in a discrete latent hierarchy. The framework combines three components: an AQ-VAE that learns a shared discrete codebook across acceleration scales, a cross-attentive transformer that predicts the next acceleration scale, and an on-policy privileged information distillation stage used for post-training.

At the core of the method is a scale-wise autoregressive prior over the latent hierarchy, where each acceleration level is predicted from all preceding levels. Concretely,
\[
Q_{32}^{\text{next}} = Q_{16},\quad
Q_{16}^{\text{next}} = Q_{8},\quad
Q_{8}^{\text{next}} = Q_{4},\quad
Q_{4}^{\text{next}} = Q_{2},\quad
Q_{2}^{\text{next}} = Q_{\text{FS}},
\]
where \text{FS} denotes the fully-sampled acquisition and the ordered acceleration factors are $\mathcal{K} = \{32,16,8,4,2\}$. This yields the factorization
\begin{equation}
p_{\theta}\big(Q_{32}, Q_{16}, \dots, Q_{\text{FS}}\big)
=
\prod_{k \in \mathcal{K}}
p_{\theta}\big(
    Q_k^{\text{next}} \mid Q_{32}, Q_{16}, \dots, Q_k
\big),
\label{eq:scalewise_likelihood}
\end{equation}
where each conditional term is parameterized by a cross-attentive transformer. Following VAR~\cite{tian2024visual}, all tokens within a scale are decoded in parallel in a single forward pass. An overview of the architecture is shown in Figure~\ref{fig:overall_var}a.

\subsection{Additive Quantized Variational Autoencoder (AQ-VAE)}
Our proposed AQ-VAE departs from the RQ-VAE \cite{lee2022autoregressive} tokenizer used in VAR~\cite{tian2024visual}, which constructs a latent hierarchy by sequentially quantizing residuals of a single latent representation. Instead, we build a natural hierarchy from inputs acquired at multiple acceleration levels, where each level contributes a different amount of information to the final representation. Highly accelerated inputs are represented with fewer tokens, while lower-acceleration inputs provide progressively richer latent detail. Their corresponding quantized maps are then fused before decoding.

We denote the continuous latent at acceleration level $k \in \{32,16,8,4,2,\text{FS}\}$ by $Z_k$ and its quantized token map by $Q_k$. Since MRI data are complex-valued, each input image is represented with real and imaginary channels. A label-informed encoder conditioned on the acceleration factor and sampling pattern (via label-dependent feature scaling and shifting) produces $Z_k$, which is quantized by nearest-neighbor $\ell_2$ lookup to obtain $Q_k$. Following~\cite{tian2024visual}, we apply a lightweight post-quantization convolution and replace the straight-through estimator with the rotation trick~\cite{fifty2025restructuring} to improve gradient flow. The refined token maps are then averaged across scales and decoded by a shared decoder for reconstruction.

\begin{figure}
\includegraphics[width=1\linewidth]{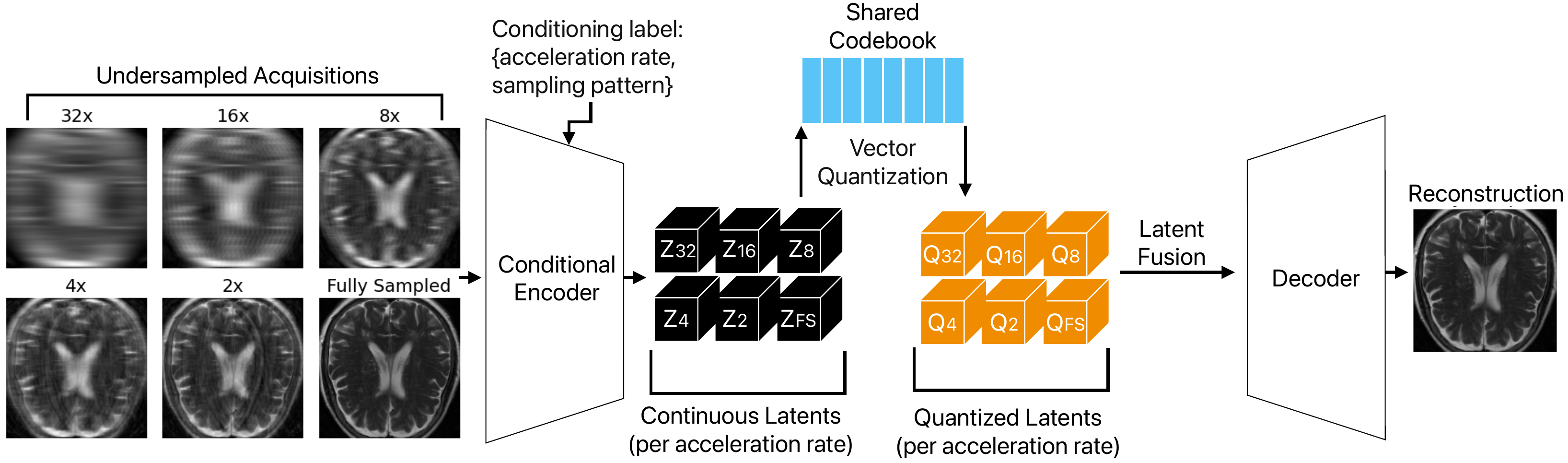}
\caption{Overview of the proposed AQ-VAE architecture.}
\label{fig:aq_vae}
\end{figure}

Compared to the hard latent hierarchy used in VAR~\cite{tian2024visual}, which progresses from a single token to a $16\times16$ grid, we adopt a lighter hierarchy better suited to MRI reconstruction. Specifically, we begin with an $11\times11$ token grid for the $32\times$ accelerated latent and increase the spatial resolution by $1\times1$ at each subsequent level until reaching a $16\times16$ grid for the fully-sampled scale. This design reflects the fact that even a $32\times$ undersampled MRI measurement retains substantially more structural information than the $1\times1$ class token used in class-conditional VAR. Moreover, because inference is performed using only the $32\times$ measurement, we aim to encode as much reliable structure as possible into $Q_{32}$. The overall architecture is illustrated in Figure~\ref{fig:aq_vae}.

We train AQ-VAE end-to-end using a combination of reconstruction, adversarial, perceptual, and commitment losses. We adopt EMA-based codebook updates and use a BiomedCLIP~\cite{zhang2023biomedclip} ViT-based discriminator, following~\cite{korkmaz2025iigalip}, as the adversarial counterpart. The overall objective is
\begin{equation}
\mathcal{L}_{\text{AQ-VAE}}
= 1.0\,\mathcal{L}_{\text{SSIM}}
+ 0.1\,\mathcal{L}_{\text{adv}}
+ 0.1\,\mathcal{L}_{\text{perc}}
+ 0.25\,\mathcal{L}_{\text{com}},
\end{equation}
where $\mathcal{L}_{\text{SSIM}}$ is the SSIM reconstruction loss, $\mathcal{L}_{\text{adv}}$ is the adversarial loss, $\mathcal{L}_{\text{perc}}$ is the perceptual loss, and $\mathcal{L}_{\text{com}}$ is the codebook commitment loss. Additional implementation details of the discriminator, encoder, decoder, and training configuration are provided in the supplementary material.

\subsection{Cross-Attentive Transformer Backbone}
\begin{figure}[h]
\centering
    \includegraphics[width=1\linewidth]{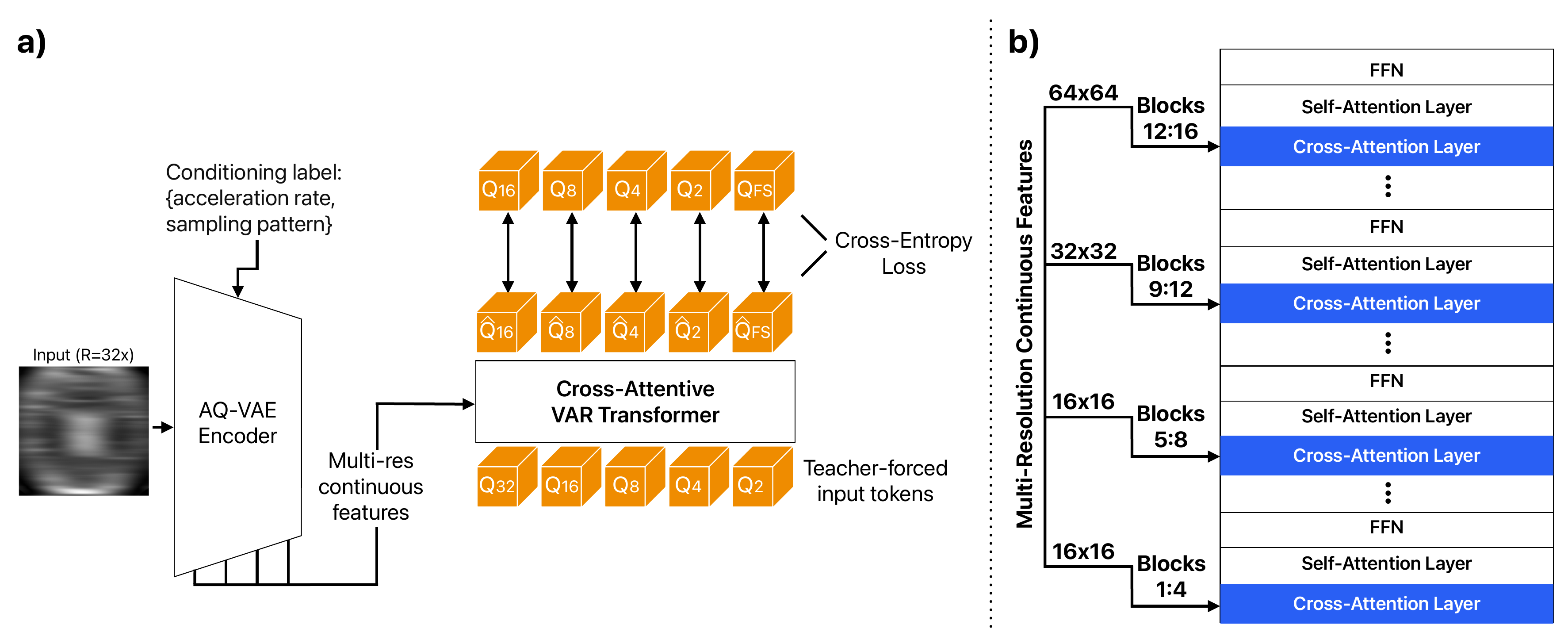}
    \caption{\textbf{(a)} Overview of the proposed cross-attentive transformer for next-acceleration-scale prediction. \textbf{(b)} The network contains 16 transformer blocks and receives encoder features at resolutions 64$\times$64, 32$\times$32, and 16$\times$16 via cross-attention, while preserving the original VAR self-attention and feed-forward components.}
    \label{fig:overall_var}
\end{figure}

VAR~\cite{tian2024visual} introduces a causal transformer for next-resolution-scale prediction that relies purely on teacher forcing to learn the data distribution. While this design is effective for natural image synthesis, we find it suboptimal for the level of fidelity required in MRI reconstruction. To provide stronger anatomical guidance, we extract multi-resolution features from our pre-trained AQ-VAE encoder at several intermediate resolutions (64$\times$64, 32$\times$32, 16$\times$16) and inject them into different layers of the modified transformer via cross-attention. Early layers receive coarse 16$\times$16 latents to enforce global structural consistency, whereas deeper layers are conditioned on progressively higher-resolution features (up to 64$\times$64), supplying detailed context for refining low-level high-frequency structures (see Figure \ref{fig:overall_var}). Our transformer is trained with teacher forcing and a cross-entropy loss to predict next-acceleration-scale token indices.

\subsection{On-Policy Privileged Information Distillation}

After training the base model, we perform an on-policy privileged information distillation step as post-training to improve rollout robustness and suppress noisy or unstable next-scale predictions, which consistently improves PSNR and SSIM across all sampling patterns and often preserves or improves perceptual quality (\cref{tab:ablation_base_vs_distilled}).

\begin{figure}[!t]
    \centering
    \includegraphics[width=1\linewidth]{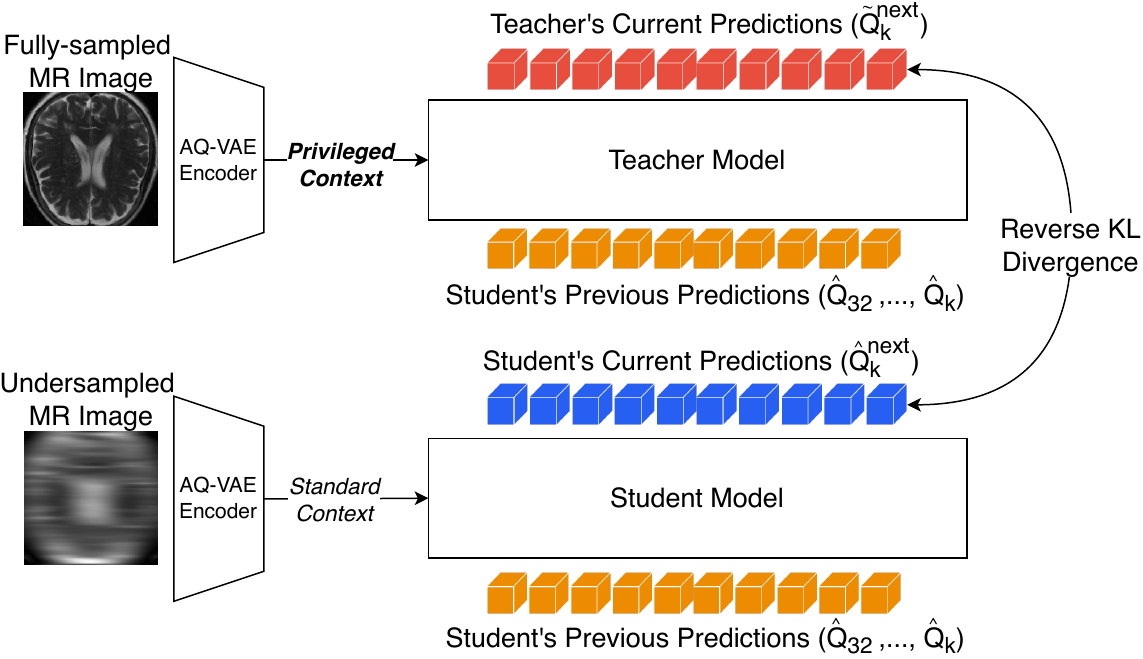}
    \caption{On-Policy Privileged Information Distillation scheme is illustrated.}
    \label{fig:distillation_scheme}
\end{figure}

In our distillation scheme, the student model autoregressively generates the latent token sequence from the undersampled MRI input, feeding each sampled token back as context for subsequent prediction steps. In parallel, a frozen teacher model observes the same partial rollout but has access to privileged information, which in our case is the fully sampled MR image, and provides a target token distribution at each scale of generation. The distillation objective minimizes the discrepancy between the student and teacher distributions, while gradients are applied only to the student (see \cref{fig:distillation_scheme}).

This formulation is on-policy rather than offline, since supervision is computed on the exact trajectories visited by the current student, including imperfect prefixes induced by its own sampling process. Consequently, the student is optimized under the same state distribution encountered at inference time. To make the teacher effective in this setting, we train it differently from the student during standard model optimization. Specifically, we expose the teacher to randomized prefix tokens, which encourages it to rely less on ideal token histories and more on the available conditioning context when predicting future tokens. This design makes the teacher better suited to guide the student when the student deviates from the ground-truth trajectory. As a result, the teacher serves not only as a privileged predictor, but also as a robust corrective signal for noisy intermediate rollouts.

Formally, let $\hat{Q}=\{\hat{Q}_{32}, \hat{Q}_{16}, \hat{Q}_{8}, \hat{Q}_{4}, \hat{Q}_{2}\}$ denote a full rollout generated by the student, and let $\hat{Q}_{\leq k}=(\hat{Q}_{32}, \hat{Q}_{16}, \dots, \hat{Q}_k)$ denote the student-generated latent history up to scale $k$. At each scale $k$, the student defines a predictive distribution $p_{\theta}(\cdot \mid \hat{Q}_{\leq k})$ over the next-acceleration-scale token vocabulary, while the frozen privileged teacher defines $p_{\phi}(\cdot \mid \hat{Q}_{\leq k}, x)$ from the same student-generated history, additionally conditioned on the fully sampled MR image $x$. We then minimize the reverse KL divergence (following \cite{ye2026policy,shenfeld2026self}) from the student to the privileged teacher across all scales, leveraging its mode-seeking behavior to favor confident teacher-supported predictions:
\begin{equation}
\mathcal{L}(\theta)
=
\mathbb{E}_{\hat{Q}\sim p_{\theta}}
\left[
\frac{1}{|\mathcal{K}|}
\sum_{k\in\mathcal{K}}
D_{\mathrm{KL}}\!\left(
p_{\theta}(\cdot \mid \hat{Q}_{\leq k})
\,\|\, 
p_{\phi}(\cdot \mid \hat{Q}_{\leq k}, x)
\right)
\right],
\label{eq:opid}
\end{equation}
where $\mathcal{K}=\{32,16,8,4,2\}$ denotes the set of acceleration scales used in the hierarchical prediction process. 

\begin{figure}[!t]
    \centering
    \includegraphics[width=.8\linewidth]{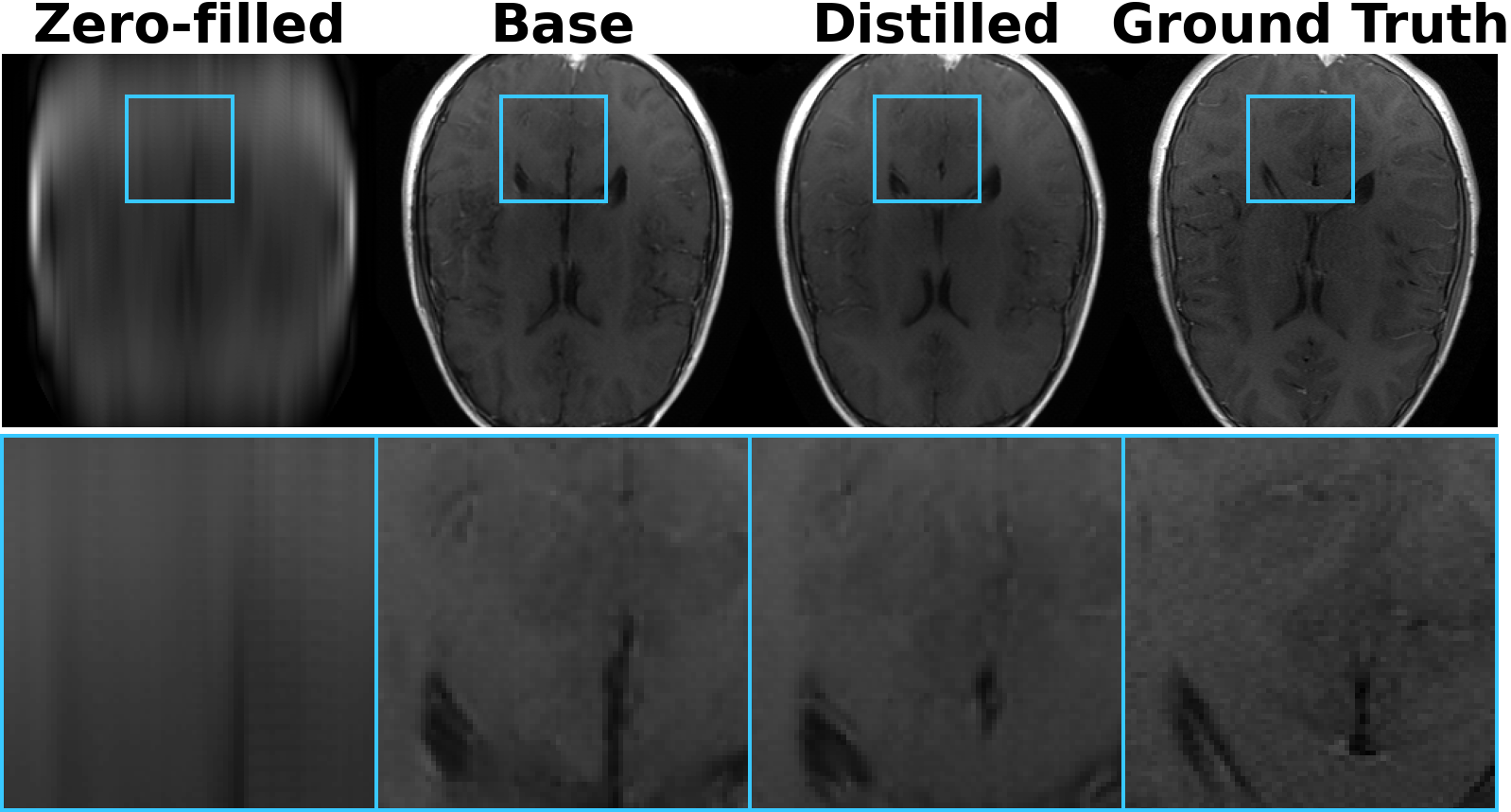}
    \caption{Distilled vs. Base model reconstructions under ES Cartesian-X undersampling. Although the overall reconstruction quality remains limited in this severe undersampling setting, Reverse KL reduces hallucinated and over-predicted details by discouraging anatomically implausible predictions.}
    \label{fig:distilled_vs_base}
\end{figure}

In particular, reverse KL strongly penalizes student probability mass assigned to outcomes that the privileged teacher considers unlikely. In our setting, this discourages unsupported next-acceleration-scale predictions and helps suppress hallucinated structures, as illustrated qualitatively in \cref{fig:distilled_vs_base}.

\subsection{Implementation Details}
The AQ-VAE uses a codebook of size 4096, latent dimension 32, and channel width 160. The cross-attentive transformer is configured with depth 16, embedding dimension 1024, 16 attention heads, MLP ratio 4.0, and drop-path rate 0.025. Base training is performed on 10 NVIDIA RTX A5000 GPUs with distributed bfloat16 training and TF32 enabled, using AdamW ($\beta_1=0.9,\beta_2=0.95$), learning rate $3.125\times10^{-5}$, and a linear decay schedule with warm-up for 250 epochs. During distillation, we train for 100 epochs on 8 NVIDIA RTX A5000 GPUs in distributed bfloat16 with global batch size 24 and a cosine learning rate schedule with base learning rate $10^{-5}$. Inference results are obtained with deterministic argmax decoding unless otherwise stated.

\section{Results}

\subsection{Evaluation Strategy}
We evaluate our method on the fastMRI \cite{fastmri} multi-coil brain benchmark under an extreme acceleration setting of $R=32$, which is particularly challenging because the acquired measurements provide only sparse constraints on the underlying image. Following our problem setup, we report results across three contrasts: T\textsubscript{1}-weighted, T\textsubscript{2}-weighted, and FLAIR. We evaluate under four undersampling patterns: Equispaced (ES) Cartesian-X, Equispaced (ES) Cartesian-Y, Radial, and 2D Gaussian Variable Density (VD). These masks induce meaningfully different reconstruction regimes rather than merely different sparsity levels. Cartesian-Y is the standard accelerated 2D Cartesian setting, while Cartesian-X provides a controlled orientation-swapped variant. 2D Gaussian VD and Radial sampling induce qualitatively different artifact patterns, allowing us to evaluate robustness across multiple corruption regimes. Additional discussion of the sampling patterns and their acquisition characteristics is provided in the supplementary material.

We compare against a broad set of competing approaches spanning several reconstruction paradigms. Specifically, we include pixel-space pure data-driven convolutional baseline UNet \cite{fastmri}, data-driven transformer-based baseline SwinUNet \cite{cao2022swin}, physics-informed unrolled methods (E2EVarnet \cite{sriram2020end} and RecurrentVarnet \cite{yiasemis2022recurrent}), diffusion-based generative reconstruction baselines (DiffuseRecon \cite{peng2022towards} and MDPG \cite{zhang2025mdpg}), and a recent strong Mamba-based physics-informed model (MambaRecon \cite{korkmaz2025mambarecon}). This set of baselines covers both purely data-driven and physics-guided reconstruction strategies, as well as modern generative approaches. 

For evaluation, we report PSNR and SSIM~\cite{wang2004image} together with feature-space perceptual metrics: unless otherwise stated, LPIPS~\cite{zhang2018unreasonable} denotes the AlexNet~\cite{krizhevsky2012imagenet}-based variant, while VGG16~\cite{simonyan2014very}-based LPIPS and DISTS~\cite{ding2020image} metrics are provided in the supplementary material. Since pixel-level metrics can favor over-smoothed reconstructions in severely ill-posed settings, perceptual measures provide a complementary view of reconstruction quality. Recent MRI studies further suggest that deep-feature similarity metrics align more closely with expert radiologist judgments than PSNR and SSIM~\cite{kastryulin2023image,adamson2025using}.

\textbf{Runtime:}
All baselines except proposed model, DiffuseRecon and MDPG are feed-forward; on an RTX A5000 our method takes $0.2441$~s/image, versus $0.8667$~s for MDPG (20 DDIM steps) and $52.27$~s for DiffuseRecon (1000 diffusion steps).

\begin{table}[h]
  \centering
  \caption{Reconstruction performance on the fastMRI (ES Cartesian-X, $R = 32$).}
  \label{tab:fastmri_brain_cartx_equispaced_x32_methods_rows}

  \small
  \renewcommand{\arraystretch}{1.08}
  \rowcolors{3}{gray!6}{white}

  \resizebox{1\textwidth}{!}{%
  \begin{tabular}{@{}l
    C{1.18cm} C{1.18cm} C{1.18cm}
    C{1.18cm} C{1.18cm} C{1.18cm}
    C{1.18cm} C{1.18cm} C{1.18cm}@{}}
    \toprule
    \textbf{Method} &
    \multicolumn{3}{c}{\textbf{PSNR} $\uparrow$} &
    \multicolumn{3}{c}{\textbf{SSIM} $\uparrow$} &
    \multicolumn{3}{c@{}}{\textbf{LPIPS} $\downarrow$} \\
    \cmidrule(lr){2-4}\cmidrule(lr){5-7}\cmidrule(l){8-10}
    &
    \textbf{T\textsubscript{1}} & \textbf{T\textsubscript{2}} & \textbf{FLAIR} &
    \textbf{T\textsubscript{1}} & \textbf{T\textsubscript{2}} & \textbf{FLAIR} &
    \textbf{T\textsubscript{1}} & \textbf{T\textsubscript{2}} & \textbf{FLAIR} \\
    \midrule

    UNet            & 19.21 & 16.94 & 17.80 & 0.60 & 0.47 & 0.46 & 0.51 & 0.58 & 0.52 \\
    SwinUNet        & 18.39 & 16.61 & 15.80 & 0.53 & 0.45 & 0.40 & 0.50 & 0.58 & 0.49 \\
    E2EVarnet      & 21.88 & 18.94 & 17.88 & 0.73 & 0.57 & 0.58 & 0.35 & 0.46 & 0.43 \\
    RecurrentVarnet & 20.53 & 18.36 & 17.27 & 0.71 & 0.55 & 0.57 & 0.38 & 0.49 & 0.45 \\
    DiffuseRecon       & 21.24 & 18.09 & 18.82 & 0.48 & 0.44 & 0.39 & 0.29 & 0.25 & 0.28 \\
    MDPG            & 21.25 & 18.25 & 17.90 & 0.63 & 0.51 & 0.48 & 0.36 & 0.47 & 0.42 \\
    MambaRecon      & 24.09 & \best{19.63} & 18.53 & \best{0.78} & \best{0.61} & \best{0.61} & 0.30 & 0.32 & 0.36 \\
    Proposed        & \best{24.61} & 19.16 & \best{21.29} & 0.76 & 0.58 & 0.60 & \best{0.16} & \best{0.24} & \best{0.22} \\

    \bottomrule
  \end{tabular}%
  }
\end{table}

\begin{table}[h]
  \centering
  \caption{Reconstruction performance on the fastMRI (Radial, $R = 32$).}
  \label{tab:fastmri_brain_radial_x32_methods_rows}

  \small
  \renewcommand{\arraystretch}{1.08}
  \rowcolors{3}{gray!6}{white}

  \resizebox{1\textwidth}{!}{%
  \begin{tabular}{@{}l
    C{1.18cm} C{1.18cm} C{1.18cm}
    C{1.18cm} C{1.18cm} C{1.18cm}
    C{1.18cm} C{1.18cm} C{1.18cm}@{}}
    \toprule
    \textbf{Method} &
    \multicolumn{3}{c}{\textbf{PSNR} $\uparrow$} &
    \multicolumn{3}{c}{\textbf{SSIM} $\uparrow$} &
    \multicolumn{3}{c@{}}{\textbf{LPIPS} $\downarrow$} \\
    \cmidrule(lr){2-4}\cmidrule(lr){5-7}\cmidrule(l){8-10}
    &
    \textbf{T\textsubscript{1}} & \textbf{T\textsubscript{2}} & \textbf{FLAIR} &
    \textbf{T\textsubscript{1}} & \textbf{T\textsubscript{2}} & \textbf{FLAIR} &
    \textbf{T\textsubscript{1}} & \textbf{T\textsubscript{2}} & \textbf{FLAIR} \\
    \midrule

    UNet            & 22.22 & 19.15 & 18.33 & 0.71 & 0.56 & 0.55 & 0.39 & 0.45 & 0.44 \\
    SwinUNet        & 20.96 & 18.45 & 17.95 & 0.62 & 0.51 & 0.48 & 0.38 & 0.42 & 0.37 \\
    E2EVarnet      & 25.27 & 21.45 & 20.40 & 0.78 & 0.66 & 0.63 & 0.27 & 0.28 & 0.31 \\
    RecurrentVarnet & 24.70 & 21.30 & 20.60 & 0.78 & 0.66 & 0.65 & 0.30 & 0.30 & 0.31 \\
    DiffuseRecon       & 26.11 & 22.97 & 22.40 & 0.60 & 0.62 & 0.51 & 0.23 & \best{0.19} & 0.24 \\
    MDPG            & 24.73 & 20.77 & 20.22 & 0.68 & 0.58 & 0.50 & 0.29 & 0.32 & 0.31 \\
    MambaRecon      & \best{27.16} & \best{24.32} & \best{23.43} & \best{0.83} & \best{0.76} & \best{0.72} & 0.23 & 0.20 & 0.23 \\
    Proposed        & 26.00 & 20.51 & 22.68 & 0.78 & 0.64 & 0.63 & \best{0.16} & 0.20 & \best{0.19} \\

    \bottomrule
  \end{tabular}%
  }

\end{table}

\subsection{Quantitative Results}
Tables~\ref{tab:fastmri_brain_cartx_equispaced_x32_methods_rows}--\ref{tab:fastmri_brain_radial_x32_methods_rows} show that our method is particularly strong in the Cartesian settings, where the induced artifact patterns are harder to resolve. Under both ES Cartesian-X and ES Cartesian-Y, it achieves the best PSNR on T\textsubscript{1} and FLAIR and the best LPIPS across all three contrasts. The gains are especially large on FLAIR, where PSNR improves from 18.53 to 21.29 for Cartesian-X and from 18.15 to 20.96 for Cartesian-Y relative to MambaRecon.

\begin{figure}[!h]
    \centering
    \includegraphics[width=1\linewidth]{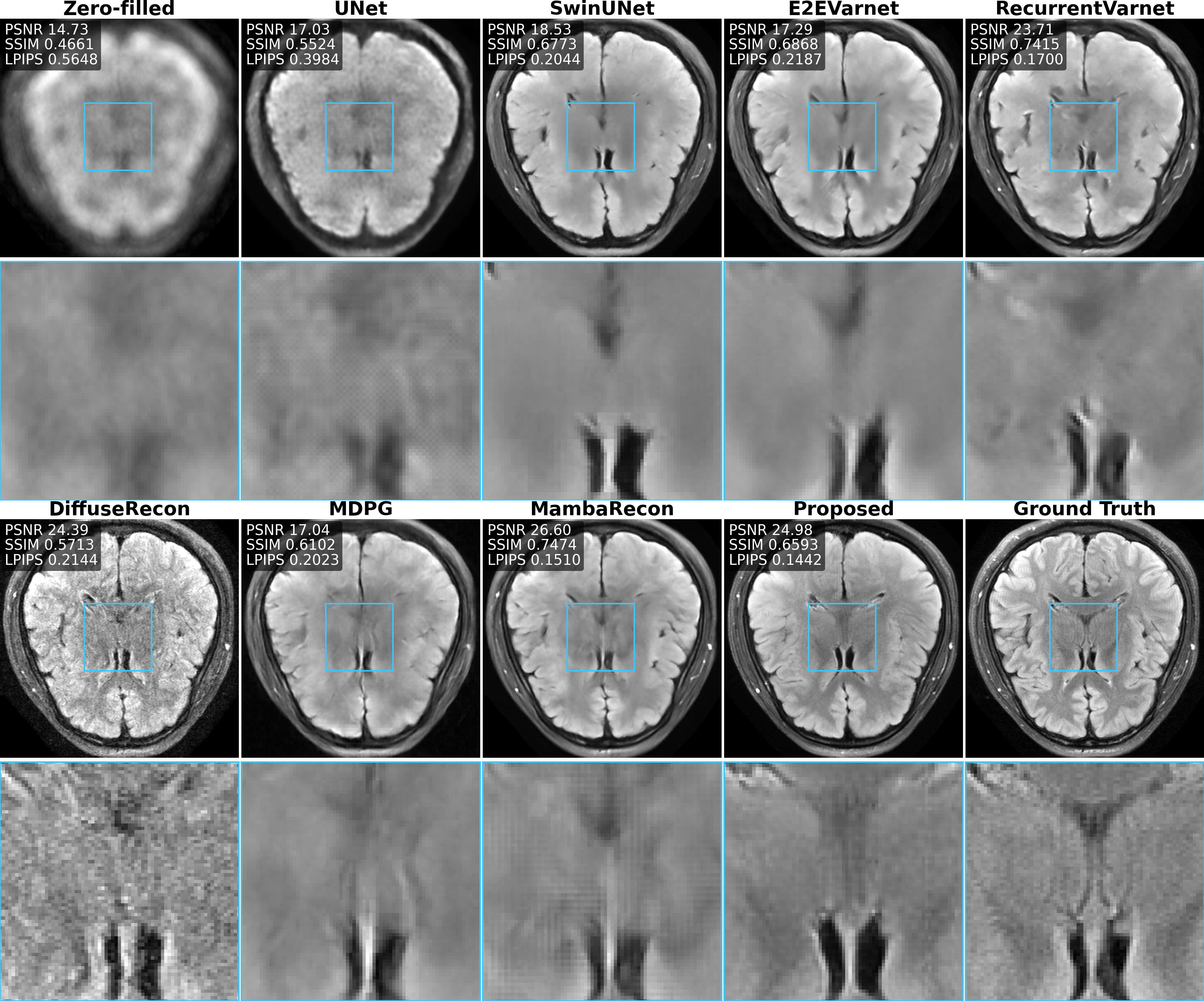}
    \caption{Qualitative comparison under Gaussian-VD undersampling. Per-image metrics are reported, and zoomed-in regions are shown below each method.}
    \label{fig:gaussian_qualitative}

\end{figure}

A different trade-off appears in the radial and Gaussian-VD settings. Here, several physics-informed continuous baselines obtain higher PSNR and SSIM, but these gains are often associated with smoother, blurrier reconstructions that suppress subtle anatomical detail and can artificially improve pixel-wise fidelity metrics. Correspondingly, our method remains strongest or highly competitive in LPIPS across contrasts in both the radial setting (Table~\ref{tab:fastmri_brain_radial_x32_methods_rows}) and the Gaussian-VD setting (Supp. Table 1). The qualitative results in Figures~\ref{fig:cartesian_x_qualitative}--\ref{fig:gaussian_qualitative} confirm this interpretation.

Overall, our method is the most consistent in feature-space perceptual quality, achieving the best average AlexNet-LPIPS, VGG-LPIPS, and DISTS for every mask type (Supp. Table 2).

\begin{figure}[!h]
    \centering
    \includegraphics[width=1\linewidth]{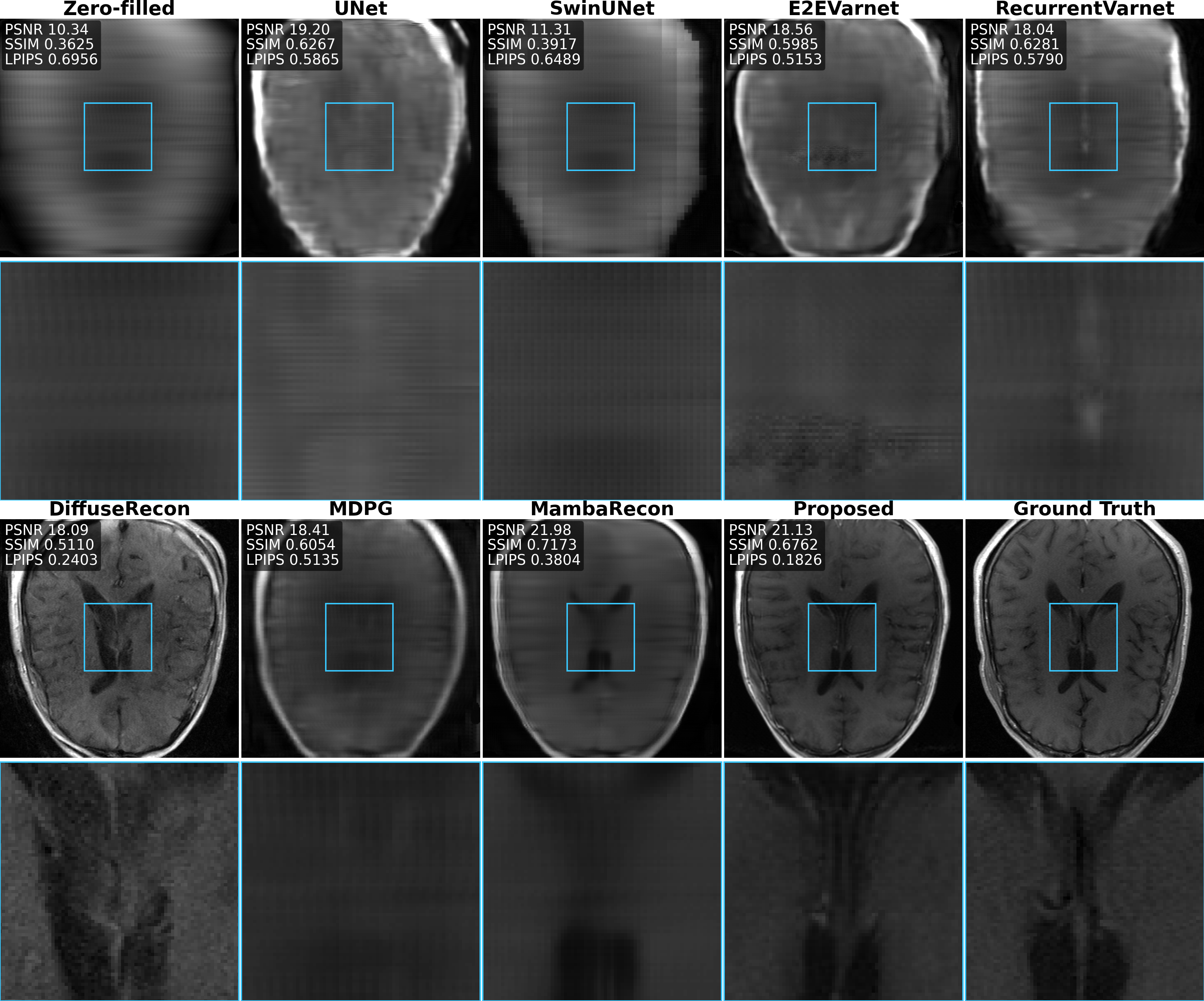}
    \caption{Qualitative comparison under ES Cartesian-Y undersampling. Per-image metrics are reported, and zoomed-in regions are shown below each method.}
    \label{fig:cartesian_y_qualitative}
\end{figure}

\begin{table}[]
  \centering
  \caption{Reconstruction performance on the fastMRI (ES Cartesian-Y, $R = 32$).}
  \label{tab:fastmri_brain_carty_equispaced_x32_methods_rows}

  \small
  \renewcommand{\arraystretch}{1.08}
  \rowcolors{3}{gray!6}{white}

  \resizebox{1\textwidth}{!}{%
  \begin{tabular}{@{}l
    C{1.18cm} C{1.18cm} C{1.18cm}
    C{1.18cm} C{1.18cm} C{1.18cm}
    C{1.18cm} C{1.18cm} C{1.18cm}@{}}
    \toprule
    \textbf{Method} &
    \multicolumn{3}{c}{\textbf{PSNR} $\uparrow$} &
    \multicolumn{3}{c}{\textbf{SSIM} $\uparrow$} &
    \multicolumn{3}{c@{}}{\textbf{LPIPS} $\downarrow$} \\
    \cmidrule(lr){2-4}\cmidrule(lr){5-7}\cmidrule(l){8-10}
    &
    \textbf{T\textsubscript{1}} & \textbf{T\textsubscript{2}} & \textbf{FLAIR} &
    \textbf{T\textsubscript{1}} & \textbf{T\textsubscript{2}} & \textbf{FLAIR} &
    \textbf{T\textsubscript{1}} & \textbf{T\textsubscript{2}} & \textbf{FLAIR} \\
    \midrule

    UNet            & 21.60 & 17.67 & 17.34 & 0.69 & 0.49 & 0.50 & 0.44 & 0.56 & 0.50 \\
    SwinUNet        & 18.02 & 14.61 & 15.45 & 0.51 & 0.36 & 0.39 & 0.57 & 0.66 & 0.53 \\
    E2EVarnet      & 21.69 & 17.71 & 17.42 & 0.70 & 0.51 & 0.54 & 0.40 & 0.51 & 0.45 \\
    RecurrentVarnet & 20.72 & 16.73 & 17.01 & 0.68 & 0.47 & 0.54 & 0.42 & 0.55 & 0.45 \\
    DiffuseRecon       & 20.22 & 16.67 & 17.81 & 0.45 & 0.37 & 0.35 & 0.31 & 0.28 & 0.31 \\
    MDPG            & 21.25 & 17.48 & 17.96 & 0.60 & 0.46 & 0.42 & 0.39 & 0.49 & 0.42 \\
    MambaRecon      & 23.40 & \best{18.89} & 18.15 & \best{0.76} & \best{0.59} & \best{0.58} & 0.32 & 0.33 & 0.34 \\
    Proposed        & \best{24.11} & 18.42 & \best{20.96} & 0.74 & 0.55 & 0.57 & \best{0.17} & \best{0.23} & \best{0.22} \\

    \bottomrule
  \end{tabular}%

  }

\end{table}

\subsection{Qualitative Results}

Figures~\ref{fig:cartesian_y_qualitative}, \ref{fig:gaussian_qualitative}, and \ref{fig:cartesian_x_qualitative}, together with additional examples in the supplementary material, reinforce the quantitative trends. Across sampling patterns, our method produces sharper and more anatomically faithful reconstructions, with cleaner tissue boundaries and better preserved fine structures. This is particularly evident in Figures~\ref{fig:gaussian_qualitative} and ~\ref{fig:cartesian_y_qualitative}, where several competing methods obtain higher per-image PSNR or SSIM but still exhibit noticeable smoothing and loss of high-frequency detail, whereas our reconstructions retain crisp structural delineation. These examples support the interpretation that the stronger pixel-level scores of several continuous baselines are driven by over-smoothed reconstructions, whereas our method better preserves local structure and perceptually meaningful detail.

\begin{figure}[h]
    \centering
    \includegraphics[width=1\linewidth]{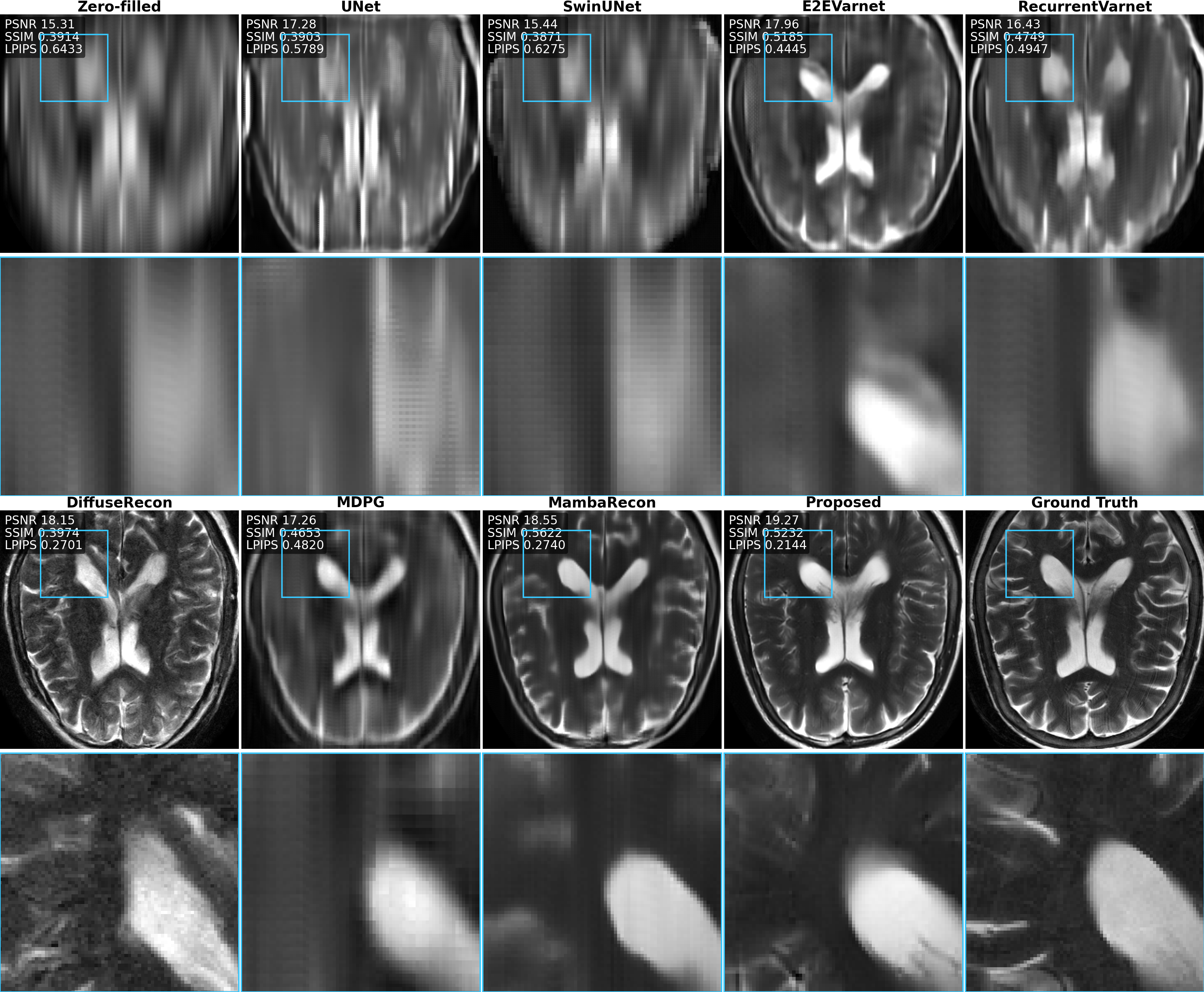}
    \caption{Qualitative comparison under ES Cartesian-X undersampling. Per-image metrics are reported, and zoomed-in regions are shown below each method.}
    \label{fig:cartesian_x_qualitative}
\end{figure}

\section{Ablation Experiments}

\subsection{Effect of On-Policy Privileged Information Distillation}
Table~\ref{tab:ablation_base_vs_distilled} shows that on-policy privileged information distillation consistently improves PSNR and SSIM across all sampling patterns and contrasts. The gains are especially clear in the more challenging Cartesian settings, where the distilled model improves both fidelity metrics for all three contrasts. LPIPS remains largely stable, with small improvements in several cases and only minor regressions in others. Overall, these results indicate that distillation improves rollout robustness and reduces anatomically implausible token predictions without materially degrading perceptual quality.

\begin{table}[h]
  \centering
  \caption{Ablations on the fastMRI validation set ($R=32$), reported as averages over all contrasts and sampling patterns (\textit{w/o} denotes \textit{without}).}
  \label{tab:ablation_components_all_masks}

  \small
  \renewcommand{\arraystretch}{1.08}
  \rowcolors{3}{gray!6}{white}

  \resizebox{1\textwidth}{!}{%
  \begin{tabular}{@{}l
    C{1.08cm} C{1.08cm} C{1.12cm}
    C{1.08cm} C{1.08cm} C{1.12cm}
    C{1.08cm} C{1.08cm} C{1.12cm}
    C{1.08cm} C{1.08cm} C{1.12cm}@{}}
    \toprule
    \textbf{Variant} &
    \multicolumn{3}{c}{\textbf{Gaussian-VD}} &
    \multicolumn{3}{c}{\textbf{ES Cartesian-X}} &
    \multicolumn{3}{c}{\textbf{ES Cartesian-Y}} &
    \multicolumn{3}{c@{}}{\textbf{Radial}} \\
    \cmidrule(lr){2-4}\cmidrule(lr){5-7}\cmidrule(lr){8-10}\cmidrule(l){11-13}
    &
    \textbf{PSNR}  & \textbf{SSIM}  & \textbf{LPIPS}  &
    \textbf{PSNR}  & \textbf{SSIM}  & \textbf{LPIPS}  &
    \textbf{PSNR}  & \textbf{SSIM}  & \textbf{LPIPS}  &
    \textbf{PSNR}  & \textbf{SSIM}  & \textbf{LPIPS}  \\
    \midrule
    w/o Cross-Attention   & 19.84 & 0.57 & 0.232 & 18.75 & 0.54 & 0.243 & 18.33 & 0.52 & 0.250 & 19.02 & 0.55 & 0.243 \\
    w/o Token Hierarchy   & 22.74 & 0.65 & 0.198 & 20.04 & 0.56 & 0.267 & 19.04 & 0.52 & 0.306 & 21.03 & 0.60 & 0.244 \\
    w/o Trainable Encoder        & 23.61 & 0.68 & 0.166 & 20.76 & 0.60 & 0.208 & 20.08 & 0.57 & 0.216 & 21.78 & 0.63 & 0.194 \\
    Base (Multinomial Sampling)        & 23.95 & 0.69 & 0.161 & 21.19 & 0.59 & 0.207 & \textbf{20.64} & 0.57 & 0.217 & 22.20 & 0.63 & 0.190 \\
    Base (Top-$p$=0.96, Top-$k$=900)   & 24.02 & 0.69 & 0.159 & \textbf{21.30} & 0.60 & 0.205 & 20.64 & 0.57 & 0.215 & 22.25 & 0.63 & 0.188 \\
    Base (Argmax Decoding) & \textbf{24.13} & \textbf{0.70} & \textbf{0.156} & 21.24 & \textbf{0.61} & \textbf{0.200} & 20.54 & \textbf{0.58} & \textbf{0.207} & \textbf{22.34} & \textbf{0.65} & \textbf{0.183} \\

    \bottomrule
  \end{tabular}%
  }

  \vspace{2pt}
  \footnotesize
  \textit{Best} values are in \textbf{bold}.

\end{table}

\subsection{Component Ablations}
Table~\ref{tab:ablation_components_all_masks} confirms that each component contributes to performance across sampling patterns. For a fair comparison, all component ablations are evaluated with the same deterministic decoding as the baseline, namely argmax token selection. Relative to Base (Argmax Decoding), the variants without Cross-Attention, without Token Hierarchy (flat $16\times16$ tokens at all scales), and without Trainable Encoder consistently reduce PSNR/SSIM and increase LPIPS, highlighting the importance of cross-attentive conditioning, a structured multi-scale token hierarchy, and jointly training the context encoder with the autoregressive predictor. We also examined the effect of stochastic decoding during autoregressive token generation. In addition to argmax selection (deterministic decoding), we evaluate multinomial sampling from the full softmax distribution and truncated sampling using top-$p$ and top-$k$ filtering. Overall, deterministic argmax decoding yields the strongest results across masks.

\begin{table}[h]
  \centering
  \caption{Reconstruction performance after on-policy privileged information distillation on the fastMRI test set ($R = 32$). Results are reported for the base model and the distilled model across different sampling patterns.}
  \label{tab:ablation_base_vs_distilled}

  \small
  \renewcommand{\arraystretch}{1.08}
  \rowcolors{3}{gray!6}{white}

  \resizebox{1\textwidth}{!}{%
  \begin{tabular}{@{}ll
    C{1.10cm} C{1.10cm} C{1.10cm}
    C{1.10cm} C{1.10cm} C{1.10cm}
    C{1.10cm} C{1.10cm} C{1.10cm}@{}}
    \toprule
    \textbf{Mask} & \textbf{Model} &
    \multicolumn{3}{c}{\textbf{PSNR} $\uparrow$} &
    \multicolumn{3}{c}{\textbf{SSIM} $\uparrow$} &
    \multicolumn{3}{c@{}}{\textbf{LPIPS} $\downarrow$} \\
    \cmidrule(lr){3-5}\cmidrule(lr){6-8}\cmidrule(l){9-11}
    & &
    \textbf{T\textsubscript{1}} & \textbf{T\textsubscript{2}} & \textbf{FLAIR} &
    \textbf{T\textsubscript{1}} & \textbf{T\textsubscript{2}} & \textbf{FLAIR} &
    \textbf{T\textsubscript{1}} & \textbf{T\textsubscript{2}} & \textbf{FLAIR} \\
    \midrule

    Gaussian-VD
      & Base      & 27.53 & 21.82 & 23.71 & 0.80 & 0.68 & 0.65 & \textbf{0.141} & 0.165 & 0.165 \\
    
      & Distilled & \textbf{27.86} & \textbf{22.29} & \textbf{24.08} & \textbf{0.81} & \textbf{0.70} & \textbf{0.67} & 0.148 & \textbf{0.161} & \textbf{0.162} \\

    \midrule

    ES Cartesian-X
      & Base      & 24.14 & 18.60 & 21.07 & 0.75 & 0.54 & 0.57 & \textbf{0.158} & \textbf{0.227} & \textbf{0.209} \\
    
      & Distilled & \textbf{24.61} & \textbf{19.16} & \textbf{21.29} & \textbf{0.76} & \textbf{0.58} & \textbf{0.60} & 0.162 & 0.241 & 0.218 \\

    \midrule

    ES Cartesian-Y
      & Base      & 23.73 & 17.84 & 20.61 & 0.73 & 0.51 & 0.54 & \textbf{0.168} & 0.231 & \textbf{0.216} \\
    
      & Distilled & \textbf{24.11} & \textbf{18.42} & \textbf{20.96} & \textbf{0.74} & \textbf{0.55} & \textbf{0.57} & 0.169 & \textbf{0.230} & \textbf{0.216} \\

    \midrule

    Radial
      & Base      & 25.51 & 19.83 & 22.34 & 0.77 & 0.60 & 0.61 & \textbf{0.156} & 0.202 & 0.190 \\
    
      & Distilled & \textbf{26.00} & \textbf{20.51} & \textbf{22.68} & \textbf{0.78} & \textbf{0.64} & \textbf{0.63} & 0.160 & \textbf{0.196} & \textbf{0.186} \\

    \bottomrule
  \end{tabular}%
  }

  \vspace{2pt}
  \footnotesize
  \textit{Best} values are in \textbf{bold}.

\end{table}

\section{Discussion and Limitations}
We study MRI reconstruction at an extreme acceleration factor of $32\times$. While this setting is useful for exposing differences between reconstruction priors, it may be too aggressive for routine clinical use because the acquired measurements can be insufficient for consistently reliable diagnostic interpretation. Therefore, our results should be interpreted as evidence of robustness in a highly ambiguous regime, rather than as a claim of clinical readiness at $32\times$ acceleration.

A key limitation of our framework is the fidelity of the discrete latent representation: the final reconstruction cannot exceed the representational precision of the learned tokenizer and codebook. This limitation is especially relevant at lower acceleration factors, where the measurements already strongly constrain the solution and continuous-valued physics-informed models can exploit the available information without passing through a quantized bottleneck. In such regimes, continuous models may remain more beneficial than discrete-token reconstruction unless substantially higher-fidelity tokenizers are developed.

At the same time, the modularity of our framework provides a natural path forward, since improved tokenizers can be used as drop-in replacements as discrete representation learning advances. More broadly, our work opens a new class of discrete-token MRI reconstruction models, bringing recent advances in autoregressive modeling to inverse problems. Discrete representations may also provide interpretability through token usage, hierarchy, and error propagation patterns that are difficult to expose in continuous-valued models, which we leave for future study. 

Finally, we show that continuous information in discrete visual autoregressive models can be used as privileged information for on-policy distillation: it is available only during training, injected through cross-attention, and used to guide the student toward rollouts that better preserve target structure.

\section{Conclusion}
We introduced a discrete autoregressive approach to accelerated MRI reconstruction, casting recovery as next-acceleration-scale prediction in a multi-scale latent token hierarchy. The framework combines an additive multi-input AQ-VAE with a cross-attentive transformer, enabling measured acquisitions to guide token prediction at every scale. A central contribution is our on-policy privileged information distillation strategy, where a teacher with access to fully sampled information supervises the student's own autoregressive rollouts, reducing exposure mismatch and discouraging unsupported structures during generation. Experiments on fastMRI under large acceleration factors show that this formulation achieves strong perceptual quality and preserves anatomically meaningful detail across diverse sampling patterns.


%
%
\clearpage
\bibliographystyle{splncs04}
\bibliography{main}
    \setcounter{figure}{0}%
    \setcounter{table}{0}%
    \renewcommand{\figurename}{Supp.~Fig.}%
    \renewcommand{\tablename}{Supp.~Table}%
    \renewcommand{\figureautorefname}{Supp.~Fig.}%
    \renewcommand{\tableautorefname}{Supp.~Table}%
    \crefname{figure}{Supp.~Fig.}{Supp.~Figs.}%
    \Crefname{figure}{Supp.~Fig.}{Supp.~Figs.}%
    \crefname{table}{Supp.~Table}{Supp.~Tables}%
    \Crefname{table}{Supp.~Table}{Supp.~Tables}%

\clearpage
\section*{Supplementary Material}
\begin{figure}[!h]
    \centering
    \includegraphics[width=1\linewidth]{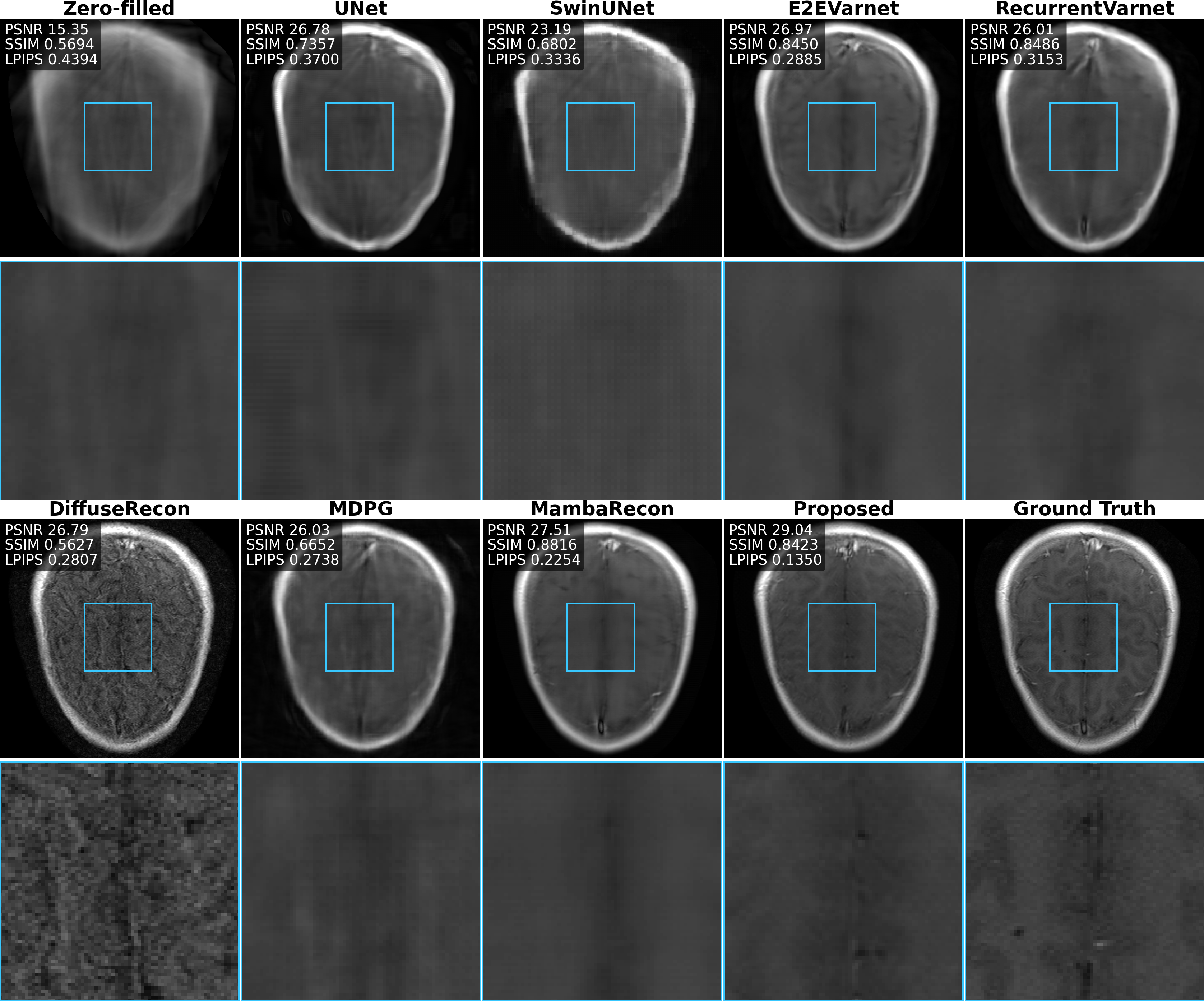}
    \caption{Qualitative comparison under Radial undersampling. Per-image metrics are reported, and zoomed-in regions are shown below each method.}
    \label{fig:radial_add}
\end{figure}

\begin{figure}[!h]
    \centering
    \includegraphics[width=1\linewidth]{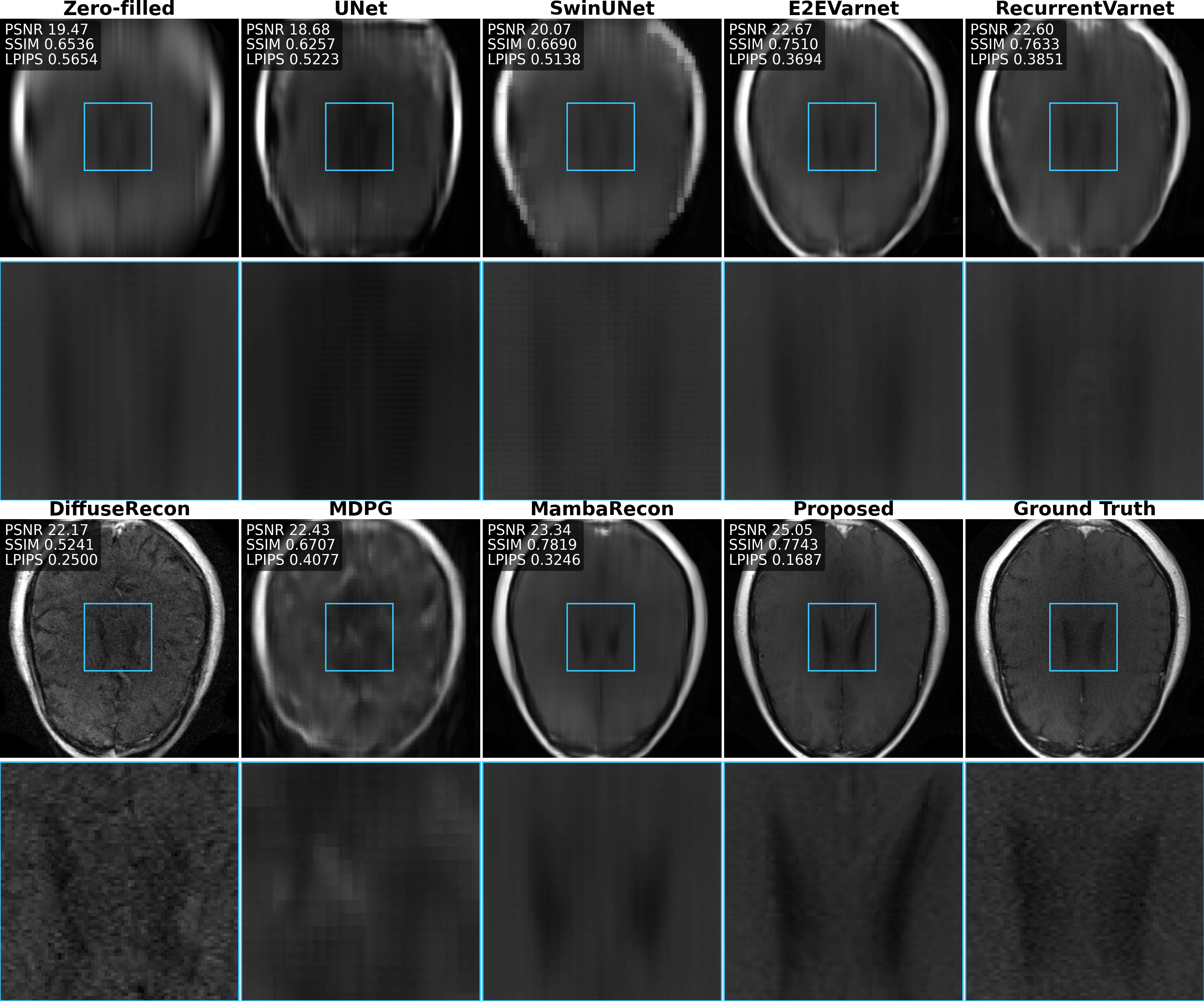}
    \caption{Qualitative comparison under ES Cartesian-X undersampling. Per-image metrics are reported, and zoomed-in regions are shown below each method.}
    \label{fig:cartesian_x_add}
\end{figure}

\begin{figure}[!h]
    \centering
    \includegraphics[width=1\linewidth]{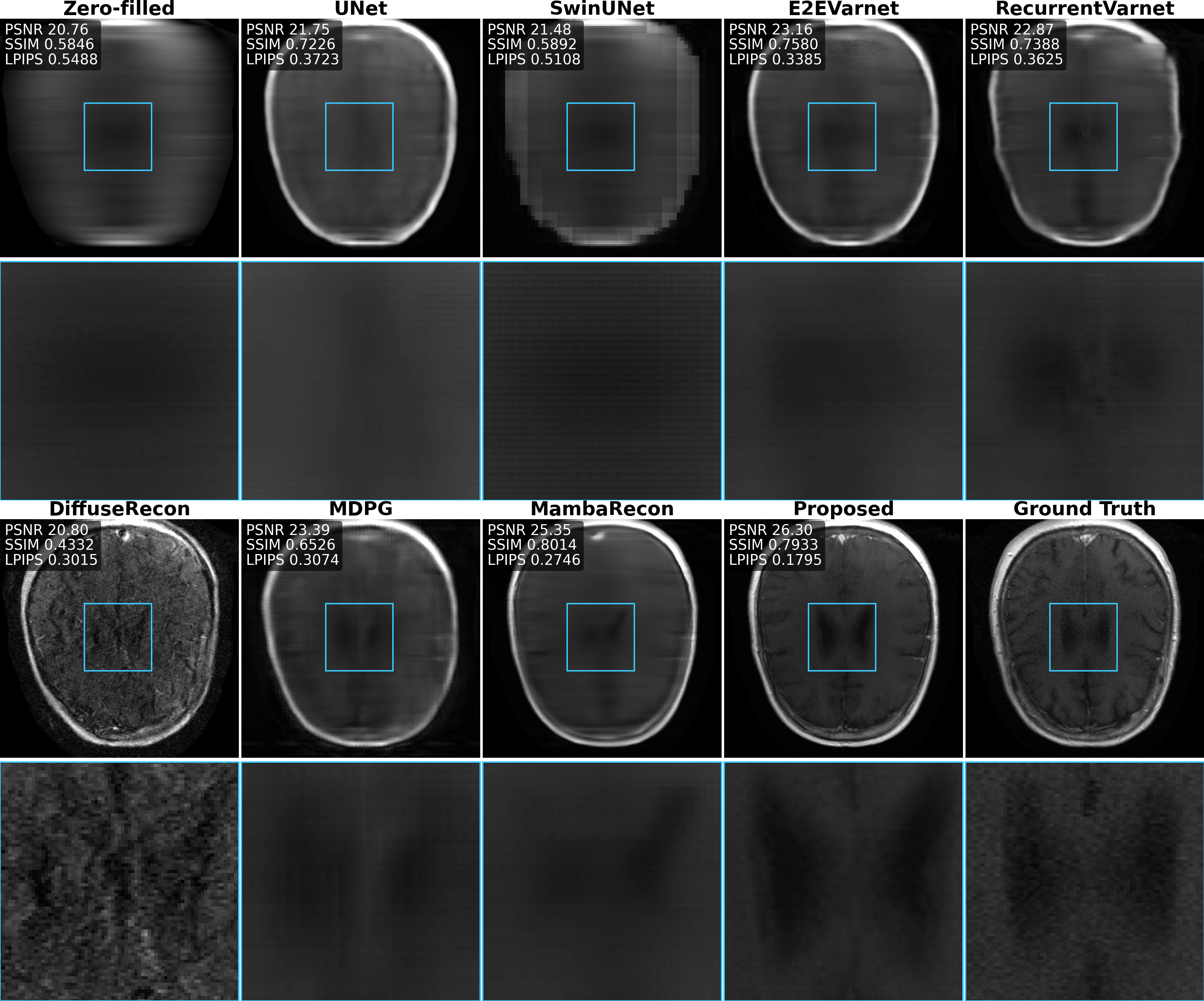}
    \caption{Qualitative comparison under ES Cartesian-Y undersampling. Per-image metrics are reported, and zoomed-in regions are shown below each method.}
    \label{fig:cartesian_y_add}
\end{figure}

\begin{figure}[!h]
    \centering
    \includegraphics[width=1\linewidth]{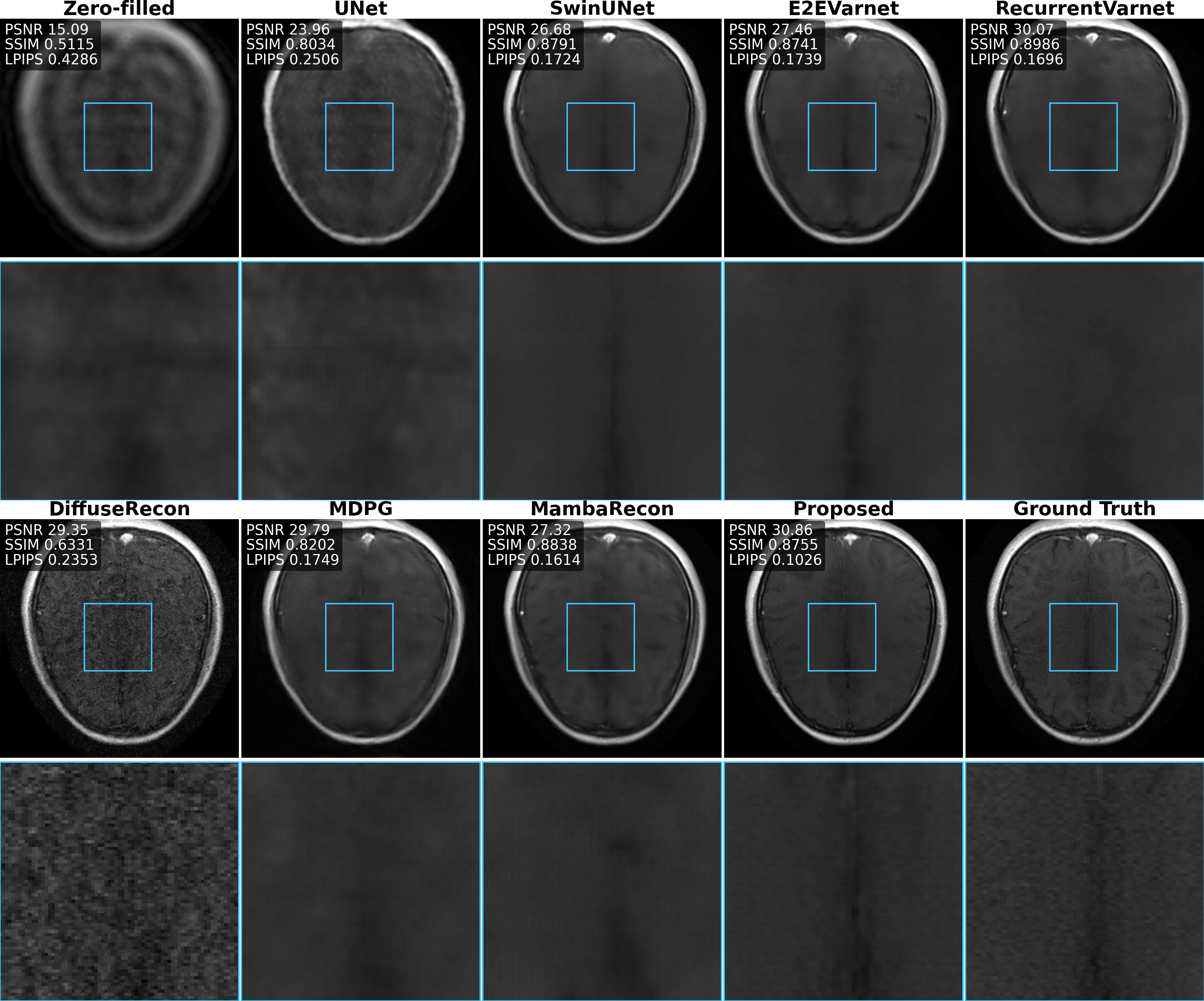}
    \caption{Qualitative comparison under Gaussian-VD undersampling. Per-image metrics are reported, and zoomed-in regions are shown below each method.}
    \label{fig:gaussian_add}
\end{figure}

This supplementary material first provides additional details on the AQ-VAE architecture and training procedure. It then discusses the undersampling patterns considered in our experiments. We conclude with extended quantitative results in \cref{tab:fastmri_brain_gaussian_vd_x32_methods_rows,tab:feature_metrics_all_masks_mean} and additional qualitative results across all mask types in \cref{fig:radial_add,fig:cartesian_x_add,fig:cartesian_y_add,fig:gaussian_add}.

\section{AQ-VAE Implementation and Training Details}
\label{sec:aqvae_details}

\textbf{Overall architecture:}
AQ-VAE is implemented as a label-conditioned, multi-scale extension of the VQ-VAE used in VAR~\cite{tian2024visual}. The model operates on a six-level reconstruction hierarchy corresponding to acceleration levels \(32\times\), \(16\times\), \(8\times\), \(4\times\), \(2\times\), and the fully sampled input. Each active input is encoded by a shared encoder, projected into a common discrete latent space, quantized using a shared codebook, and then fused before decoding. In our final configuration, the latent dimensionality is \(32\), the codebook size is \(4096\), and the base channel width is \(160\). The encoder and decoder channel multipliers are \([1,1,2,2,4]\), and the model uses \(2\) residual blocks per stage. We initialize the backbone from the pretrained VQ-VAE of VAR~\cite{tian2024visual}.

\textbf{Conditional encoder:}
We adapt the VQ-VAE encoder of VAR~\cite{tian2024visual} to produce acquisition-aware latent representations for MRI reconstruction. Specifically, AQ-VAE conditions the encoder on both the acceleration factor and the sampling pattern so that feature extraction can vary with the acquisition setting. To achieve this, we replace the standard ResNet blocks in the original VAR encoder with label-informed FiLM-style modulation~\cite{perez2018film}, where each block applies learned feature-wise scaling and shifting based on the conditioning label. This allows the encoder to adjust its internal feature statistics according to the acceleration level and mask type. Unless they belong to the newly introduced conditioning pathway, encoder weights are initialized from the pretrained VAR checkpoint~\cite{tian2024visual}.

\textbf{Decoder:}
We retain the original VAR decoder~\cite{tian2024visual} and initialize it from the same pretrained checkpoint. Since the decoder is designed for three-channel inputs, whereas MRI data are represented using two channels corresponding to the real and imaginary components, we append a dummy third channel to preserve compatibility with the pretrained architecture without modifying its structure. After multi-scale fusion in latent space, the averaged fused latent is decoded into the final complex-valued reconstruction.

\textbf{Multi-scale tokenization and fusion:}
Let \(Z_{32}, Z_{16}, Z_{8}, Z_{4}, Z_{2}\), and \(Z_{\mathrm{FS}}\) denote the continuous latents extracted from the \(32\times\), \(16\times\), \(8\times\), \(4\times\), \(2\times\), and fully sampled inputs, respectively, and let \(Q_{32}, Q_{16}, Q_{8}, Q_{4}, Q_{2}\), and \(Q_{\mathrm{FS}}\) denote their corresponding quantized codes. Each active scale is resized to its own token resolution before vector quantization and then mapped back to a common latent resolution before fusion. The token grids are asymmetric across scales, with $11,\,12,\,13,\,14,\,15,\,16$
tokens per side for the \(32\times\), \(16\times\), \(8\times\), \(4\times\), \(2\times\), and fully sampled inputs, respectively. Thus, coarser inputs are quantized using smaller token grids, while finer inputs are represented at progressively higher token resolutions. After quantization, each scale passes through a scale-indexed residual quantization transform, and the resulting latent contributions are summed and divided by the number of active scales. The residual quantization strength is set to \(0.5\), and no latent normalization is applied.

\textbf{Discriminator:}
For adversarial training, we use the ViT-based visual encoder of BiomedCLIP~\cite{zhang2023biomedclip} as a medical domain feature extractor and attach lightweight projected discriminator heads to multiple intermediate transformer layers. Concretely, following~\cite{korkmaz2025iigalip}, we use the activations from transformer blocks \(2\), \(5\), \(8\), and \(11\), yielding four feature levels that capture increasingly high-level representations. This design follows prior findings in~\cite{sauer2023stylegan,korkmaz2025iigalip}, where layer-wise projected heads were shown to provide a strong and stable perceptual discriminator. The resulting discriminator operates on BiomedCLIP ViT-B/32 features from these four layers and outputs a single real-versus-fake prediction.

\textbf{Codebook learning and quantization training:}
We employ exponential moving average (EMA)-based codebook updates~\cite{razavi2019generating}, and therefore do not rely on the explicit codebook regression objective used in the original VAR VQ-VAE~\cite{tian2024visual}. In addition, we replace the straight-through estimator with the rotation trick~\cite{fifty2025restructuring}, which improves gradient propagation from the reconstruction objective to the encoder while better preserving angular relationships between encoder outputs and codebook vectors. In practice, this improves codebook utilization and increases perplexity across scales. The codebook is reinitialized at the start of training, while dead-code reinitialization during training is disabled.

\textbf{Commitment objective:}
Let \(k \in \{32,16,8,4,2,\mathrm{FS}\}\) denote the acceleration level, with continuous latent \(Z_k\) and quantized token map \(Q_k\) as defined in the main text. For each active scale \(k\), we define the commitment objective as
\[
\mathcal{L}_{\mathrm{com}}^{(k)}
=
\beta \cdot
\operatorname{MSE}\!\left(
Z_k,\, \operatorname{sg}[Q_k]
\right),
\qquad
\beta = 0.25,
\]
where \(\operatorname{sg}[\cdot]\) is the stop-gradient operator. Because the codebook entries are updated via EMA, this loss encourages the encoder outputs \(Z_k\) to stay close to their assigned codebook vectors without directly optimizing the codebook through gradients. Averaging over the all scales \(\mathcal{K}\), we obtain
\[
\mathcal{L}_{\mathrm{com}}
=
\frac{1}{|\mathcal{K}|}
\sum_{k \in \mathcal{K}}
\mathcal{L}_{\mathrm{com}}^{(k)}.
\]

\textbf{Reconstruction loss:}
We use a combined SSIM and perceptual reconstruction objective. Concretely, the model is optimized using an SSIM-based loss on the magnitude image together with an LPIPS perceptual term computed on rescaled magnitude images. The reconstruction weight is fixed to \(\lambda_{\mathrm{recon}}=1.0\), and the perceptual weight is \(\lambda_{\mathrm{perceptual}}=0.1\). No curriculum schedule is used, and the loss weights remain fixed throughout training.

\textbf{Adversarial objective:}
Let \(\phi(\cdot)\) denote the BiomedCLIP-based feature extractor~\cite{zhang2023biomedclip}, and let \(D(\cdot)\) denote the corresponding projected discriminator. We use a least-squares GAN objective \cite{mao2017least}. The discriminator loss is
\[
\mathcal{L}_D
=
\frac{1}{2}
\left[
\operatorname{MSE}\!\left(D(\phi(x)),\, 1\right)
+
\operatorname{MSE}\!\left(D(\phi(\hat{x})_{\mathrm{detach}}),\, 0\right)
\right],
\]
where \(x\) is the ground-truth image and \(\hat{x}\) is the reconstruction. The generator-side adversarial term is
\[
\mathcal{L}_{\mathrm{adv}}
=
\operatorname{MSE}\!\left(D(\phi(\hat{x})),\, 1\right).
\]
We set the adversarial weight to \(\lambda_{\mathrm{adv}}=0.1\).

\textbf{Optimization and training setup:}
Training is performed for \(100\) epochs with batch size \(16\). We use AdamW for both generator and discriminator with learning rates \(10^{-4}\) and weight decay \(10^{-4}\). Gradient clipping is set to \(5.0\) for the generator and \(1.0\) for the discriminator. We do not use mixed precision and do not apply gradient accumulation beyond a single step. Learning-rate scheduling is enabled with \(10\) warmup epochs, \(10\) cosine-decay epochs, and a minimum learning rate of \(10^{-6}\). The best checkpoint is selected according to validation SSIM.

\clearpage
\newpage
\section{Undersampling Patterns}
\label{sec:undersampling_patterns}

\begin{figure}[!h]
    \centering
    \includegraphics[width=1\linewidth]{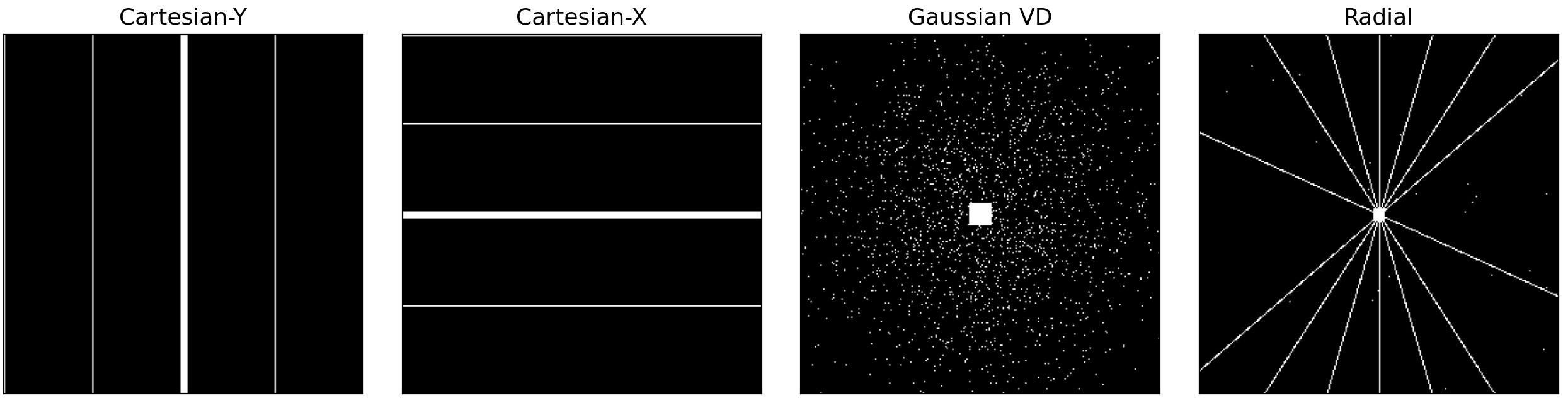}
    \caption{Representative examples of the four undersampling patterns used in our experiments at $32\times$ acceleration.}
    \label{fig:mask_types}
\end{figure}

To complement the discussion in the main text, we provide additional details on the four undersampling masks used throughout our experiments: \textbf{Equispaced (ES) Cartesian-X}, \textbf{Equispaced (ES) Cartesian-Y}, \textbf{Gaussian Variable Density (VD)}, and \textbf{Radial}. Although all masks are matched to the same nominal acceleration factor, they do not define equivalent reconstruction problems. Instead, they impose different measurement geometries, preserve different portions of k-space, and produce qualitatively different artifact patterns in image space. Representative examples of the four mask types are shown in ~\cref{fig:mask_types}.

\textbf{ES Cartesian-Y:}
Equispaced Cartesian-Y corresponds to the conventional accelerated 2D Cartesian MRI setting, where undersampling is performed along the phase-encoding direction. This is the standard acquisition regime most commonly associated with accelerated MRI and therefore serves as the primary reference setting in our evaluation. From a Fourier perspective, removing regularly spaced measurements along this axis leads to highly structured aliasing, with image content folding along the corresponding spatial dimension. Under extreme acceleration, the corruption is therefore coherent and directional rather than noise-like. Reconstruction in this setting requires the model to resolve substantial ambiguity caused by overlapping anatomical content, making Equispaced Cartesian-Y the canonical test bed for accelerated MRI reconstruction.

\textbf{ES Cartesian-X:}
Equispaced Cartesian-X is obtained by swapping the undersampling orientation of Equispaced Cartesian-Y, producing a controlled transposed variant of the standard Cartesian setup. This mask preserves the overall Cartesian nature of the acquisition while changing the directional structure of the missing information. Consequently, the artifact pattern is reoriented relative to Equispaced Cartesian-Y, even though the nominal acceleration remains the same. We include Equispaced Cartesian-X to test whether performance is robust to the orientation of the corruption pattern rather than being overly tied to the standard phase-encoding direction.

\textbf{Gaussian VD:}
The Gaussian VD mask is a 2D non-uniform sampling pattern that retains k-space measurements with higher probability near the center and lower probability toward the periphery. This follows the common MRI principle that low-frequency components carry most of the global structural and contrast information, whereas higher frequencies capture fine anatomical detail. Compared with equispaced Cartesian masks, Gaussian VD produces a qualitatively different reconstruction regime. Because the sampling is variable-density rather than periodic, the resulting artifacts are generally less dominated by coherent directional folding and instead appear more diffuse and spatially distributed. At the same time, prioritizing the center of k-space tends to preserve coarse image structure, shifting the main challenge toward recovering sharpness, boundaries, and subtle high-frequency details under severe undersampling.

\textbf{Radial:}
The Radial mask departs most strongly from the Cartesian setting. Instead of sampling along horizontal or vertical Cartesian lines, radial sampling acquires measurements along spokes passing through the center of k-space at different angles. Even under aggressive undersampling, this trajectory repeatedly covers the low-frequency central region while only sparsely sampling outer k-space directions. As a result, radial undersampling produces a distinct artifact profile, classically characterized by streaking rather than directional fold-over aliasing. These streaks are globally distributed and arise from limited angular coverage, yielding a corruption pattern that differs substantially from the Cartesian masks. We include Radial sampling to evaluate whether the reconstruction framework remains robust when the artifact structure is governed primarily by trajectory geometry rather than axis-aligned omission of k-space lines. In our implementation, to match the target \mbox{$32\times$} acceleration precisely, we supplement the radial spokes with a very small number of randomly sampled k-space points.

\clearpage
\newpage
\section{Additional Qualitative Results}

\clearpage
\newpage
\section{Additional Quantitative Results}

\begin{table}[!h]
  \centering
  \caption{Reconstruction performance on the fastMRI (Gaussian-VD, $R = 32$).}
  \label{tab:fastmri_brain_gaussian_vd_x32_methods_rows}

  \small
  \renewcommand{\arraystretch}{1.08}
  \rowcolors{3}{gray!6}{white}

  \resizebox{.85\textwidth}{!}{%
  \begin{tabular}{@{}l
    C{1.18cm} C{1.18cm} C{1.18cm}
    C{1.18cm} C{1.18cm} C{1.18cm}
    C{1.18cm} C{1.18cm} C{1.18cm}@{}}
    \toprule
    \textbf{Method} &
    \multicolumn{3}{c}{\textbf{PSNR} $\uparrow$} &
    \multicolumn{3}{c}{\textbf{SSIM} $\uparrow$} &
    \multicolumn{3}{c@{}}{\textbf{LPIPS} $\downarrow$} \\
    \cmidrule(lr){2-4}\cmidrule(lr){5-7}\cmidrule(l){8-10}
    &
    \textbf{T\textsubscript{1}} & \textbf{T\textsubscript{2}} & \textbf{FLAIR} &
    \textbf{T\textsubscript{1}} & \textbf{T\textsubscript{2}} & \textbf{FLAIR} &
    \textbf{T\textsubscript{1}} & \textbf{T\textsubscript{2}} & \textbf{FLAIR} \\
    \midrule

    UNet            & 24.67 & 20.51 & 19.72 & 0.74 & 0.59 & 0.59 & 0.35 & 0.43 & 0.40 \\
    SwinUNet        & 27.32 & 23.16 & 22.33 & 0.83 & 0.73 & 0.71 & 0.25 & 0.20 & 0.24 \\
    E2E-Varnet      & 28.41 & 24.48 & 22.75 & 0.83 & 0.75 & 0.71 & 0.23 & 0.20 & 0.24 \\
    RecurrentVarnet & \best{29.76} & 25.64 & \best{25.40} & \best{0.85} & 0.78 & \best{0.75} & 0.22 & 0.16 & 0.20 \\
    DiffRecon       & 28.89 & \best{26.36} & 24.92 & 0.66 & 0.71 & 0.59 & 0.20 & \best{0.14} & 0.20 \\
    MDPG            & 27.69 & 23.17 & 22.77 & 0.77 & 0.67 & 0.60 & 0.24 & 0.20 & 0.23 \\
    MambaRecon      & 27.90 & 25.56 & 24.99 & 0.84 & \best{0.78} & 0.74 & 0.20 & 0.15 & 0.18 \\
    Proposed        & 27.86 & 22.29 & 24.08 & 0.81 & 0.70 & 0.67 & \best{0.15} & 0.16 & \best{0.16} \\

    \bottomrule
  \end{tabular}%
  }
\end{table}

\begin{table}[!h]
  \centering
  \caption{Mean feature-space perceptual metrics across T\textsubscript{1}, T\textsubscript{2}, and FLAIR on the fastMRI brain dataset ($R = 32$). Lower is better for all metrics.}
  \label{tab:feature_metrics_all_masks_mean}

  \small
  \renewcommand{\arraystretch}{1.08}
  \rowcolors{3}{gray!6}{white}

  \resizebox{\textwidth}{!}{%
  \begin{tabular}{@{}l
    C{1.22cm} C{1.22cm} C{1.12cm}
    C{1.22cm} C{1.22cm} C{1.12cm}
    C{1.22cm} C{1.22cm} C{1.12cm}
    C{1.22cm} C{1.22cm} C{1.12cm}@{}}
    \toprule
    \textbf{Method} &
    \multicolumn{3}{c}{\textbf{Cartesian-X} } &
    \multicolumn{3}{c}{\textbf{Cartesian-Y}} &
    \multicolumn{3}{c}{\textbf{Gaussian-VD} } &
    \multicolumn{3}{c@{}}{\textbf{Radial} } \\
    \cmidrule(lr){2-4}\cmidrule(lr){5-7}\cmidrule(lr){8-10}\cmidrule(l){11-13}
    &
    \textbf{Alex} & \textbf{Vgg} & \textbf{Dists} &
    \textbf{Alex} & \textbf{VGG} & \textbf{DISTS} &
    \textbf{Alex} & \textbf{VGG} & \textbf{DISTS} &
    \textbf{Alex} & \textbf{VGG} & \textbf{DISTS} \\
    \midrule
    UNet            & 0.53 & 0.51 & 0.38 & 0.50 & 0.49 & 0.33 & 0.39 & 0.44 & 0.30 & 0.43 & 0.46 & 0.31 \\
    SwinUNet        & 0.53 & 0.53 & 0.45 & 0.59 & 0.56 & 0.48 & 0.23 & 0.34 & 0.21 & 0.39 & 0.48 & 0.35 \\
    E2EVarnet      & 0.41 & 0.43 & 0.30 & 0.45 & 0.46 & 0.31 & 0.22 & 0.32 & 0.20 & 0.29 & 0.37 & 0.23 \\
    RecurrentVarnet & 0.43 & 0.46 & 0.33 & 0.48 & 0.49 & 0.33 & 0.20 & 0.31 & 0.21 & 0.31 & 0.39 & 0.26 \\
    DiffuseRecon       & 0.27 & 0.37 & 0.21 & 0.30 & 0.39 & 0.23 & 0.18 & 0.28 & 0.18 & 0.22 & 0.31 & 0.19 \\
    MDPG            & 0.42 & 0.47 & 0.33 & 0.44 & 0.48 & 0.32 & 0.22 & 0.37 & 0.23 & 0.31 & 0.43 & 0.27 \\
    MambaRecon      & 0.33 & 0.39 & 0.27 & 0.33 & 0.40 & 0.28 & 0.18 & 0.29 & 0.18 & 0.22 & 0.30 & 0.20 \\
    Proposed        & \textbf{0.21} & \textbf{0.30} & \textbf{0.19} & \textbf{0.20} & \textbf{0.31} & \textbf{0.18} & \textbf{0.16} & \textbf{0.26} & \textbf{0.16} & \textbf{0.18} & \textbf{0.28} & \textbf{0.17} \\
    \bottomrule
  \end{tabular}%
  }
\end{table}

\end{document}